\documentclass [12pt,twoside]{article}           
\usepackage[left,modulo]{lineno}
\usepackage{setspace}

\countdef\arxiv=201

\ifnum\arxiv=0 \input cpreb.tex \fi

\ifnum\arxiv=1

\usepackage{epsfig,times,lscape}
\usepackage[usenames]{color}

    \ifnum\count102=1

    \fi

    \ifnum\count102=2
\fi

\newcommand{\nc}{\newcommand}

     \ifnum\count101=1
\nc{\qI}[1]{\section{{#1}}}
\nc{\qA}[1]{\subsection{{#1}}}
\nc{\qun}[1]{\subsubsection{{#1}}}
\nc{\qa}[1]{\paragraph{{#1}}}

\def\qpar{\vskip 2mm plus 0.2mm minus 0.2mm}
\def\qL{\hfill \break}
     \fi 

      \ifnum\count101=2
 \nc{\qI}[1]{\parindent=0mm \vskip 8mm 
{\centerline{\LARGE \color{red}#1}}\vskip 3mm}
\nc{\qA}[1]{\vskip 2.5mm \noindent 
{{\bf\large\color{blue}  #1}} \vskip 1mm \parindent=0mm}
 \nc{\qun}[1]{\vskip 1mm \noindent {\sl #1 }\quad }

\def\qL{\hfill \break}
\def\qpar{\vskip 2mm plus 0.2mm minus 0.2mm}

      \fi

\def\qth{\vrule height 12pt depth 0pt width 0pt}
\def\qtb{\vrule height 0pt depth 5pt width 0pt}
\def\tvi{\vrule height 12pt depth 5pt width 0pt}

\nc{\qfoot}[1]{\footnote{{#1}}}

      \ifnum\count101=1
\def\qbu{\hfill \par \hskip 6mm $ \bullet $ \hskip 2mm}
\def\qee#1{\hfill \par \hskip 6mm (#1) \hskip 2 mm}
      \fi
      \ifnum\count101=2
\def\qbu{\hfill \par \hskip 4mm $ \bullet $ \hskip 2mm}
\def\qee#1{\hfill \par \hskip 4mm (#1) \hskip 2 mm}
      \fi

\def\qparr{ \vskip 1.0mm plus 0.2mm minus 0.2mm \hangindent=10mm
\hangafter=1}

     \ifnum\count101=1 
 \def\qdec#1{\parindent=0mm\par {\leftskip=2cm {#1} \par}}
     \fi
     \ifnum\count101=2
  \def\qdec#1{\parindent=0mm \par {\leftskip=1cm {#1} \par}}
  
  \def\qcitb#1{\noindent \hbox to 102mm{\hfill \small #1} \vskip 1mm}
      \fi

 \def\qpages#1{\count102=0{\loop\advance\count102 by 1
 \null \vfill\eject \ifnum\count102<#1 \repeat}}

\def\tvi{\vrule height 12pt depth 5pt width 0pt}

\def\qth{\vrule height 12pt depth 0pt width 0pt}
\def\qtb{\vrule height 0pt depth 5pt width 0pt}

\def\qv{\vskip 0.1mm plus 0.05mm minus 0.05mm}
\def\qhu{\hskip 0.6mm}
\def\qhv{\hskip 3mm}

\def\qhw{\hskip 1.5mm}
\def\qleg#1#2#3{\noindent {\bf \small #1\qhw}{\small #2\qhw}{\it \small #3}\qv }
\fi

\begin{document}
\thispagestyle{empty}



\markboth{{\sl \hfill  \hfill \protect\phantom{3}}}
        {{\protect\phantom{3}\sl \hfill  \hfill}}

\color{yellow} 
\hrule height 20mm depth 10mm width 170mm 
\color{black}
\vskip -2.2cm 

 \centerline{\bf \Large Effect of marital status on death rates.}
\vskip 2mm
 \centerline{\bf \Large Part 2: Transient mortality spikes}
\vskip 16mm
\centerline{\large 
Peter Richmond$ ^1 $ and Bertrand M. Roehner$ ^2 $
}

\vskip 8mm
\normalsize
{\bf Abstract}\quad 

We examine what happens in a population when it experiences
an abrupt change in surrounding conditions. 
Several cases of such ``abrupt transitions'' for both physical and
living social systems are analyzed from which it can be seen that all
share a common pattern. 
First, a steep rising death rate followed by a much slower relaxation process 
during which the death rate decreases as a power law
(with an exponent close to 0.7). This leads us to propose a
general principle which can
be summarized as follows: {\it ``ANY abrupt change in living
conditions generates a mortality spike which acts as a 
kind of selection process''}. 
This we term the {\it Transient Shock} conjecture. It
provides a qualitative model which leads to testable predictions.
For example, marriage certainly brings about a major change
in environmental and social conditions and according to our
conjecture one would expect a mortality spike in
the months following marriage. At first sight this may seem an
unlikely proposition but we demonstrate (by three different methods)
that even here the existence of mortality spikes is supported by solid
empirical evidence.

\vskip 2mm
\centerline{\it Version of 19 August 2015. Comments are welcome.}

\vskip 1mm
{\small Key-words: death rate, marital status, widowhood, young
widowers, response function, transient behavior.
\vskip 2mm

{\normalsize 
1: School of Physics, Trinity College Dublin, Ireland.
Email: peter\_richmond@ymail.com \qL
2: Institute for Theoretical and High Energy Physics (LPTHE),
University Pierre and Marie Curie, Paris, France. 
Email: roehner@lpthe.jussieu.fr
}

\vfill\eject

\large

\qI{Introduction}

\qA{Merits and shortcomings of the death rate ratio approach}

The present paper is a continuation of Richmond and Roehner
(2015) which for the sake of brevity, will be referred to as ``Paper 1''. 
\qpar
In paper 1 was shown that in all age groups 
the death rates $ d_s, d_w, d_d $ of single, widowed or
divorced persons were higher than the death rates $ d_m $ 
of married persons.  The ratios $ r_s=d_s/d_m,\ r_w=d_w/d_m,\
r_d=d_d/d_m $ were called {\it death rate ratios} 
(or simply death ratios)
with respect to married persons. \qL
The important point we note here is that this is not
a small effect. Most death ratios are
higher than two and they become as high as 5 
for young widowers.
Gompertz's law allows conversion of death ratios 
into what may be called ``equivalent aging''.
According to Gompertz's law, after the age of 30, the death
rate doubles by 10 years of age. Thus, if the death ratio of
a widower of age 30 is equal to 3, Gompertz’s law implies that 
widowhood will push up his death rate to that of a married
man about 16 years older.
\qpar
From a statistical perspective the death ratios are 
convenient and effective variables. They are convenient
because they can be easily computed from the death rates
by age and marital status. They are effective 
in the sense that they remove the massive effect of aging
on death rates. Whereas the death rates of both married and
unmarried persons increase exponentially with age,
their ratios remain bounded within fairly narrow intervals.
\qpar

However, the death ratios are also
fairly opaque variables which do not tell us anything 
about  the actual mechanism of the FB effect. 
This is because the death ratios provide only a static picture. 
They do not
say how death rates are affected 
in the course of time by a change for example in marital status.
In other words, they do not tell us how such a transition
should be described at the level of a cohort of persons.
It is only through a longitudinal analysis in which one
follows a cohort in the course of time
that one can gain an insight into what really happens.

\qA{Life as an equilibrium state in a domain of the parameter space}

What is life? This is a question that may be answered in many
ways. For the purpose of the present paper it will be
sufficient to observe that it is an equilibrium in which
a number of parameters remain confined within fairly narrow
limits. For instance the body temperature should remain 
within 35 and 45 degree Celsius. The domain of the parameter
space which is compatible with life may be referred to as the
life envelope%
\qfoot{A similar expression is used in aviation.
The flight envelope or service envelope of an aircraft
designates the domain of the flight parameters in which 
the aircraft should remain.}%
.
Three observations are in order
in relation with the present study.
\qbu In contrast with the case of body temperature for which
the limits are rather strict, for many other parameters the
limits are fairly elastic. Consider
the concentration of hemoglobin in blood. Whereas 
the references values (for women) are 12--15g
per deciliter of blood, life
remains possible even for levels as low as 4g/dL. In addition,
such boundaries are  also subject to inter-individual variations.
The notion of frailty which is often used in relation
with elderly
persons can be seen as a global contraction of the life envelope.
\qbu Medicine focuses on biological parameters. Yet, for human
beings social factors are also very important. 
This is shown very clearly by the fact that
(as seen in Paper 1) death rates of non-married persons are two or 
three times higher than death rates of married persons.
In other words, a major change in familial and social ties
can drastically affect the life expectancy of people. 
Because, up to now, 
we have no means for measuring the strength of social interactions,
it is impossible to define a range of reference values for such
variables, however one should keep in mind the existence of such
limits.
\qbu Usually, in the process leading to death it is not just
one but
several parameters which go beyond their reference intervals. 
One reason for this is that the parameters are not independent.
This collective effect can be summarized by the notion 
of ``will to live''. Testimonies suggest that often 
the ``will to live'' disappears one or two months prior
to the actual occurrence of death.
Although this notion has probably
an objective significance, we recognize
that (so far) it has not been measured and quantified. 
The transient death spikes analyzed in the present paper may
be seen as an attempt to define this notion quantitatively.
A mortality spike reflects a change in the will to live for
the simple reason that it covers a time interval (usually
a few months) which is too short for new diseases to 
fully develop. 
In other words, the persons who die were positioned near
the limits of their life envelope.
\qpar

The previous comments give rise to several
unknown questions and open issues.
A useful guide to these comes from the observation that there
is a close connection between physical systems and systems of
living organisms. This is explained in the next section.
\qpar

The paper is in three parts.
In the next section we develop a system theory perspective
which will give us a simplified framework for the analysis of
systems of living organisms. 
It will be seen that,
the most visible effect of a state transition in a 
population is often the occurrence of a transient mortality spike.
In this way,
simply by shedding the items that are unsuitable in the new
situation, the system adapts to the environment change.
We then analyze several examples
of sharp transitions characterized by such transient mortality spikes.
This leads us to the formulation of the Transient Shock conjecture.
Finally, we test a key prediction of this conjecture 
according to which one should expect a mortality spike in
the months following marriages. 
For that purpose we explore the death rate of newly married
persons in the months following their marriage.
The challenge is to see whether there is a mortality spike or not.

\qI{System theory perspective}

In order to get a broader
understanding we will adopt a system theory point of view
which means that we will examine several systems during their
transition from a state $ A $ to a state $ B $. Establishing 
connections between systems which, at first sight, seem very
different yet nevertheless follow the same law
is typical of the approach used in physics. 
An apple and the Moon may be very different in appearance, yet 
as shown by Newton, they are ruled by the same gravitational law.
It will be seen that the transient
behavior of physical systems is fairly
similar to what is observed in the transitions occurring in
human systems. For instance when a collection of 
VLSI (Very Large Scale Integrated)
semi-conductor chips are put in operation, there is first
a period of high failure rate. 
This time of excess failure rate which may last for a few months
is commonly referred to as being an ``infant mortality'' phase
by reliability engineers.
\qpar
Evidence taken from physical as well as human systems will lead us
to the conjecture that in a transition $ A \rightarrow B $ 
there are usually {\it two} steps and not just one.
\qee{1} First, there is a short-term transition shock which
results in an upsurge of failures.
\qee{2} Secondly, there is a long-term change in the failure rate as the
system gets adapted to its new state. 
\qpar

%
\begin{figure}[htb]
\centerline{\psfig{width=12cm,figure=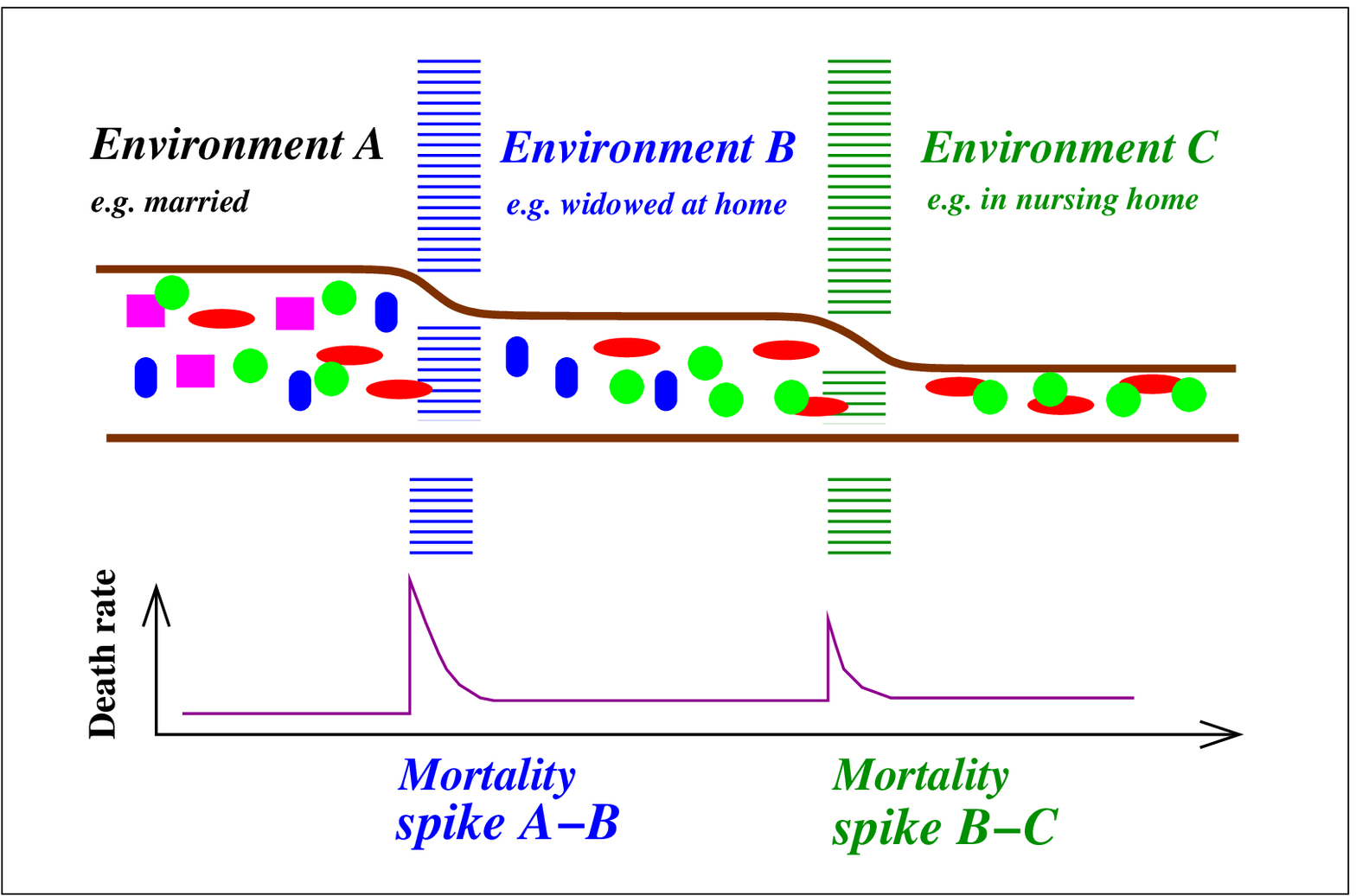}}
\qleg{Fig.\qhu 1\qhv Successive transition shocks.}
{The figure shows a system of items which goes successively
through three environments. Although all of same type, these items
are not completely identical. Their variability is represented
by their different shapes: circles, ellipses or squares.
At each transition the items which are not adapted to the new
environment ``die'' and are eliminated. 
As examples one can give the following cases.
(i) A set of lightbulbs going through three states: $ A $=turned off,
$ B $=turned on, 60 volts, $ C $=turned on, 220 volts. 
(ii) New born babies:
$ A $=pre-natal, $ B $=first day of post-natal life, $ C $=first year of life.
In the transition $ A \rightarrow B $, the deaths due to 
``defects'' (premature birth,
malformations, congenital debility, injuries at birth) represent
94\% of the total deaths whereas in the transition 
$ B \rightarrow C $ these
causes represent only 30\% of the deaths (Mortality Statistics 1910,
p. 154)
(iii) a cohort of 
persons going through three states: $ A $=married, $ B $=widowed at
home, $ C $=widowed in nursing home. }
{\it }
\end{figure}
%
\begin{figure}[htb]
\centerline{\psfig{width=8cm,figure=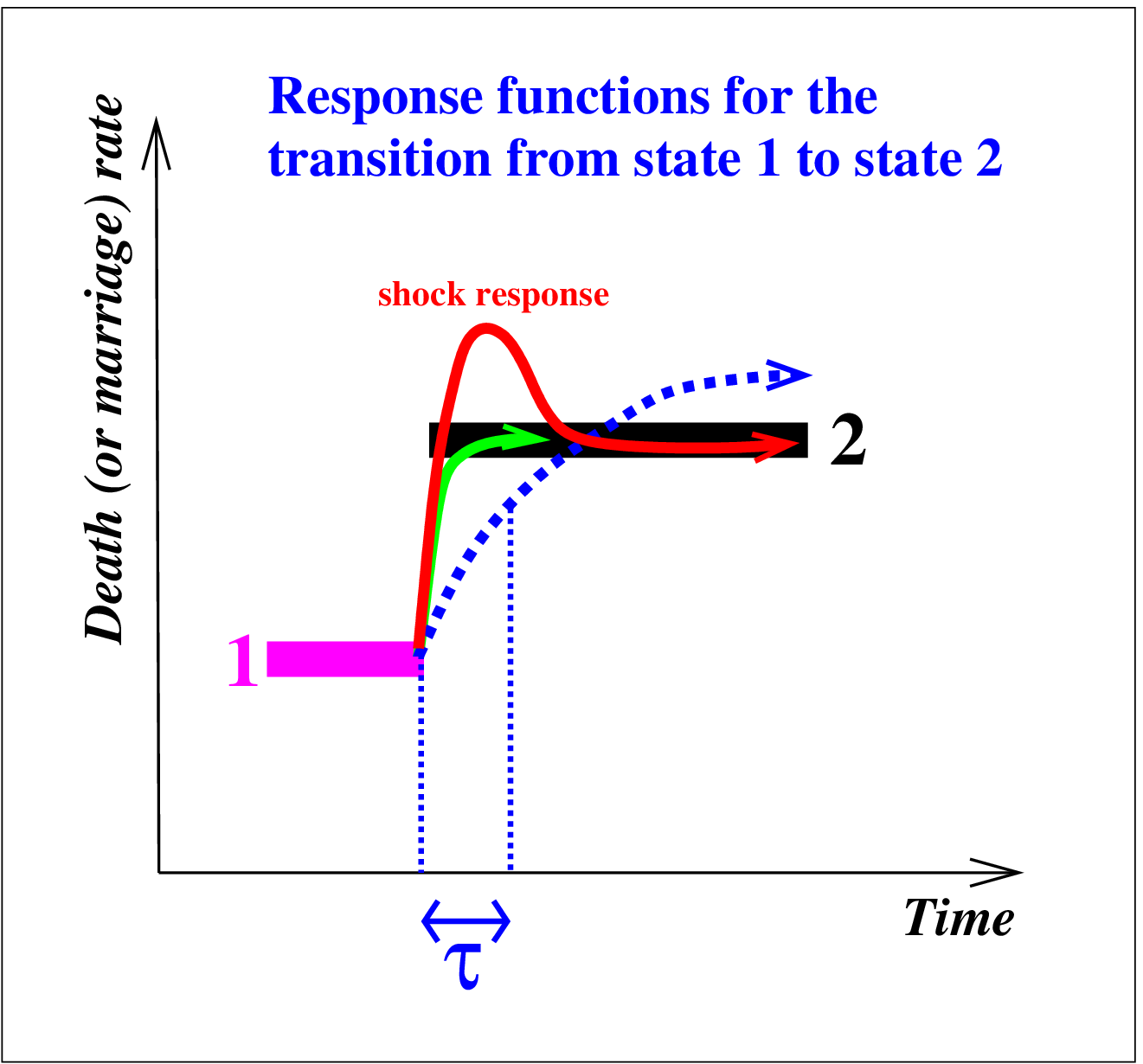}}
\qleg{Fig.\qhu 2a\qhv  Transient response functions.}
{State 1 and 2 characterize two marital situations.
One can distinguish two kinds of responses: those with overshoot
(red curve) and those without overshoot (green or blue curves).
The later converge steadily toward their stationary value.
The time constant $ \tau $ defines the duration of the 
transient response. Thus, the transition defined by the green curve has a 
shorter time constant than the one corresponding to the
blue curve.}
{}
\end{figure}

\begin{figure}[htb]
\centerline{\psfig{width=16cm,figure=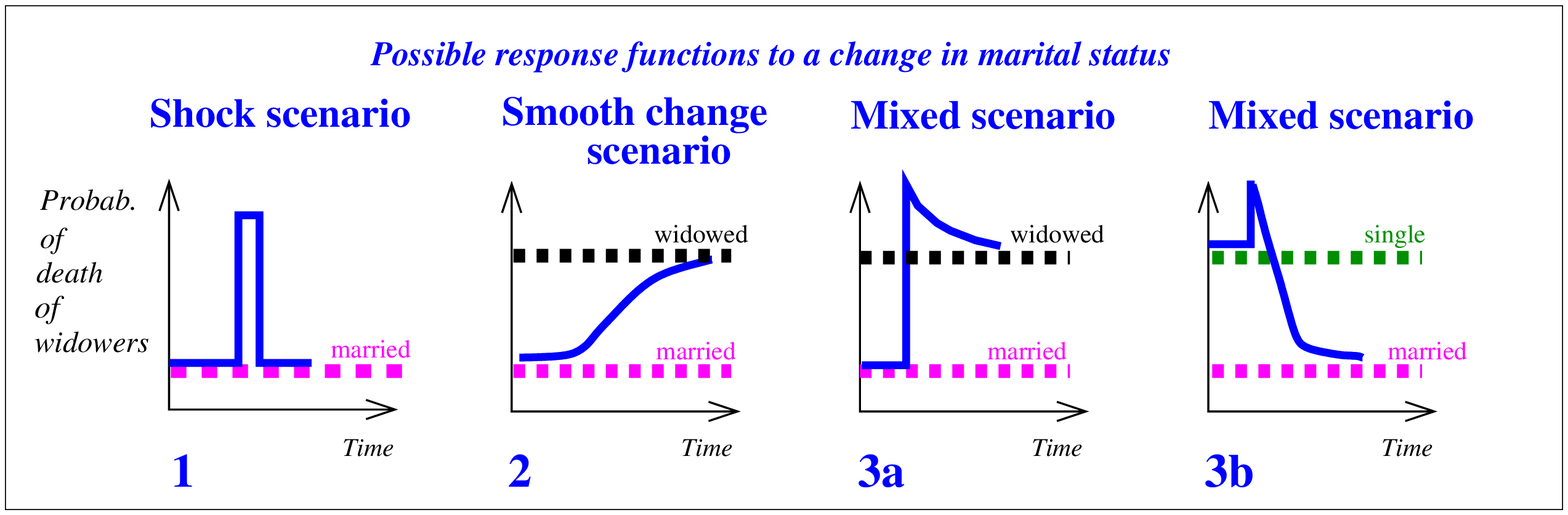}}
\qleg{Fig. \qhu 2b\qhv  Different transition scenarios.}
{The pure shock scenario 1 has no slow relaxation phase.
The smooth change scenario 2 has no steep upgoing phase.
The mixed scenarios 3a and 3b have a steep upgoing phase
as well as a slow relaxation phase. 3a is observed in the transition
married-to-widowed; 3b is observed  
in the transition single-to-married.}
{}
\end{figure}
The expression ``gets adapted'' may raise a question. How
can there be an adaptation for physical devices? Is adaptation
not a feature that is specific to living systems?
In the expression ``the system gets adapted'' the word system
does not refer to a single item but to a sample of items,
for instance it will designate a batch of, say
1,000, semi-conductor chips. Although these chips look identical
and were manufactured through the same production process,
they are in fact slightly different. Some have small defects
which will drastically reduce their life-time. Little by
little the chips with defects will fail and be eliminated.
As a result, the set of ``surviving'' chips will 
globally gain a smaller failure rate than the initial sample.
It is in this sense that the whole
set of chips becomes better adapted to its new state.
\qpar

In passing we note that
Darwin's theory of ``survival of the fittest'' also applies
to this process? Although a little reflection quickly shows that in the 
present case
the catch phrase ``survival of the fittest'' is nothing
but a tautology. For the chips
the very fact of surviving also
defines the ``fittest'' chips. For
living organisms, one may imagine that generation after generation
there is an adaptation to a changing environment. Yet, the
example of the sample of chips shows that for selection
and adaptation to take place there is no need
for any transformation at individual level. The only requirement
is that in the initial sample there is a dispersion of
some of its characteristics. 

In the following subsections we give several examples
of parallels between physical and biological systems.

\qA{Infant mortality for lightbulbs}

Everybody has observed that most often lightbulbs fail
when they are turned on. Why is this so?

When an incandescent light bulb is switched on, one has the impression
that the light goes on instantaneously. Yet, physicists 
know that natural phenomena are never instantaneous.
In the case of lightbulbs there
are two very different time scales.
\qbu An {\it electrical time lag} is due to 
the self inductance $ L $ of the
filament. $ L $ may be small but it is not zero. If one assumes
that $ L $ is of the order of one micro-Henry and the resistance of
the filament at room temperature about one Ohm, the time constant
$ L/R $ of the light bulb will be of the order of one microsecond.
Although small, such a time lag can be easily measured with
an oscilloscope. Modern digital oscilloscopes have 
sweep speeds ranging from picoseconds per division to seconds 
per division.
\qbu In order to emit light a tungsten filament must reach
a temperature of about $ 2,500 $ degree Celsius. Needless to say, this
takes much longer than a few microseconds. A reasonable order of
magnitude is about $ 100 $ms. As the filament becomes hotter, the
resistivity of tungsten increases strongly;
it gets multiplied by a factor of about $ 10 $.
This means that for at least $ 10 $ms  
the ``inrush'' current
will be some 10 times greater than the standard operating current.
In other words, there will be an overshoot phenomenon in which
the current will greatly exceed its steady-state value.
That is why light bulbs often fail immediately after
being turned on. In more expensive lamps this problem is overcome
through a preheating 
phase during which the voltage will increase progressively.\qL
Incidentally, this example shows that overshooting does
not necessarily imply that the response of the system is ruled by a second
order linear differential equation%
\qfoot{In this case overshooting occurs for critical- or under-damping
 and the response involves oscillations
which converge toward the steady-state limit.}%
.
Overshooting proves that the system is {\it not} ruled
by a {\it linear} first order  differential equation; however
a  {\it nonlinear} first order equation is clearly not excluded.
\qpar
This example shows that by observing the
response function of a system one can identify and better
understand various phenomena which take place in the
system. We seek to do the same kind of analysis for
groups of persons moving from one marital state to another.

\qA{Bathtub curves}

In a general way, 
the physical items for which the notions of failure and 
life-time have a significance are items that cannot be repaired.
Examples are light bulbs, fluorescent lamps, electronic
chips, hard drives%
\qfoot{On the contrary, for items which can be repaired,
such as shoes or cars, the notion of life-time has no real meaning.}%
.
For such items it is customary to distinguish 3 successive periods.
\qbu After the item has been turned on, there is an infant
mortality period characterized by a decreasing failure rate.
\qbu It is followed by a ``useful life'' marked by a failure
rate that is low and relatively stable.
\qbu Finally comes a wear-out period during which the failure rate
increases.
\qpar
Because of its shape with two upgoing sides, this curve is commonly
called a bathtub curve. The graph of the mortality rate
of many living organisms has a similar shape (see the graph below
for human mortality rates). 

\qA{Bathtub curve for hard drives}

A physical case of bathtub curve can be observed for hard-drives. 
The following observation was made in 2013 and 
relies on a sample of several thousands 
hard-drives in use at the BackBlaze cloud storage company.
\qbu The so-called infant mortality period lasted 1.5 years 
and is characterized
by an annual failure rate of 5\%.
\qbu The useful life lasted also 1.5 years and it has a failure rate
of 1.4\%.
\qbu Finally, the wear-out period started on average 3 years 
after operation
was started. It is characterized by a failure rate of 12\%. During
this phase the number of ``surviving hard-drives'' will progressively
decrease at a fixed rate (which means an exponential decrease).
After 4 years in operation 
about 80\% of the devices were still working.

\qA{Bathtub curve for human populations}

The three phases considered in reliability control
are clearly visible for human mortality rates particularly
if a logarithmic scale is used for the time axis.

%
\begin{figure}[htb]
\centerline{\psfig{width=16cm,figure=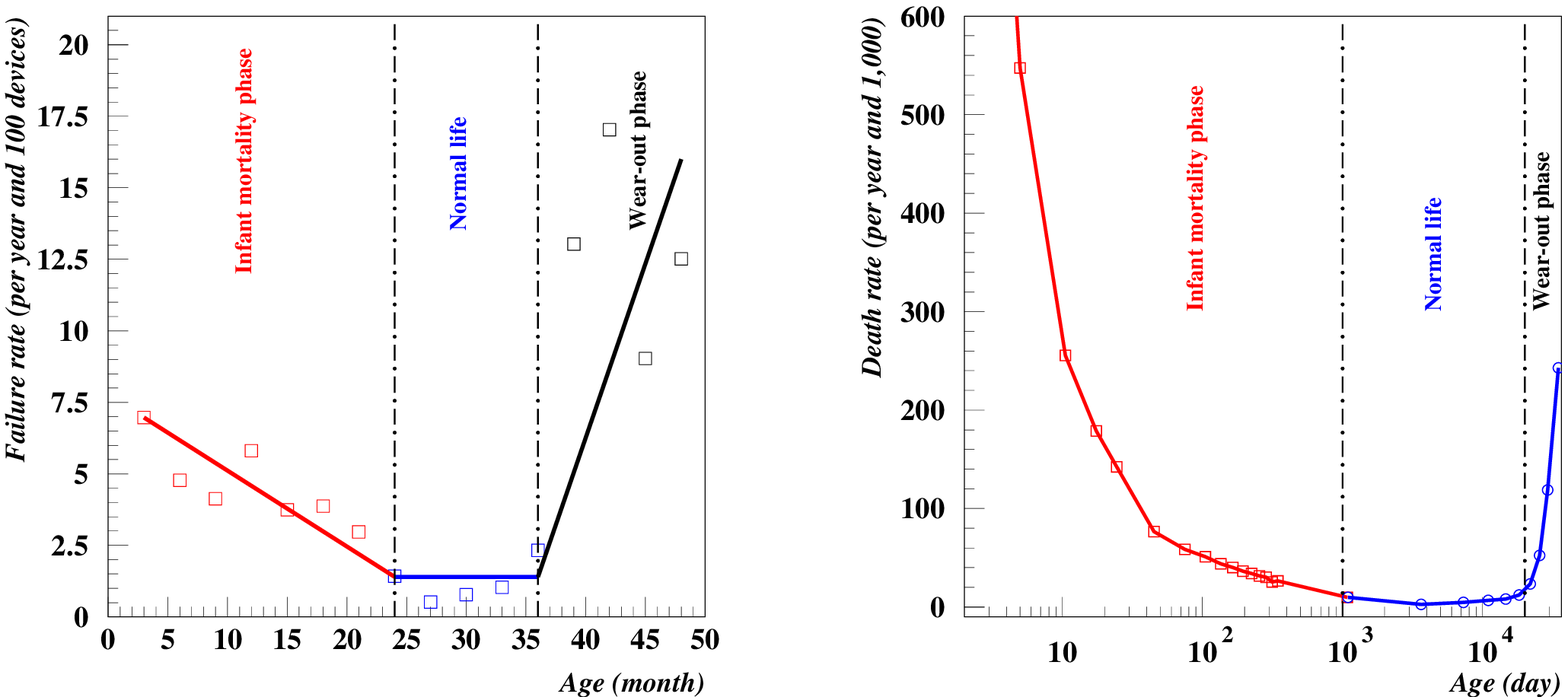}}
\qleg{Fig.\qhu 3\qhv Two examples of bathtub curves.}
{Left: Annual failure rate of hard drives used by the cloud storage
company Backblaze. The data are based on a set of several thousand
drives operating without interruption. The large dispersion
is partly due to the fact that the set included drives from
different manufacturers.
Right: Human mortality rates, male and female, USA, 1923.
The decrease with age
follows a power law: $ y\sim 1/t^{\alpha},\
\alpha=0.88\pm 0.05 $ (with a confidence probability level of 0.95),
whereas the increase is an exponential with a doubling time of about
10 years. In 1910 the exponent of the power law was equal to
$ 0.65\pm 0.04 $.}
{\it Sources: Hard drives: Beach (2013). Human mortality:
(i) Infant mortality: Linder and Grove (1947, p. 574-575).
(ii) Adult mortality: Linder and Grove (1947, p. 150).}
\end{figure}

\qI{Analysis of transient mortality spikes}

\qA{Birth as first statistical evidence of transient mortality spikes}


The transition from pre-natal to post-natal life is probably
the most dramatic change in human existence. Therefore,
if our conjecture is correct, one would expect a transient
mortality spike of great amplitude. Currently, in the most
advanced countries the infant mortality rate (i.e. mortality
during the first year) is of the order of 2 per 1,000 live births
which is about 10 times higher than the death rate in the 
age interval 5-14. As shown in the graph below,
the death rates reached
in the days and weeks immediately after birth are much
higher than the infant mortality rate.

\begin{figure}[htb]
\vskip -3mm
\centerline{\psfig{width=8cm,figure=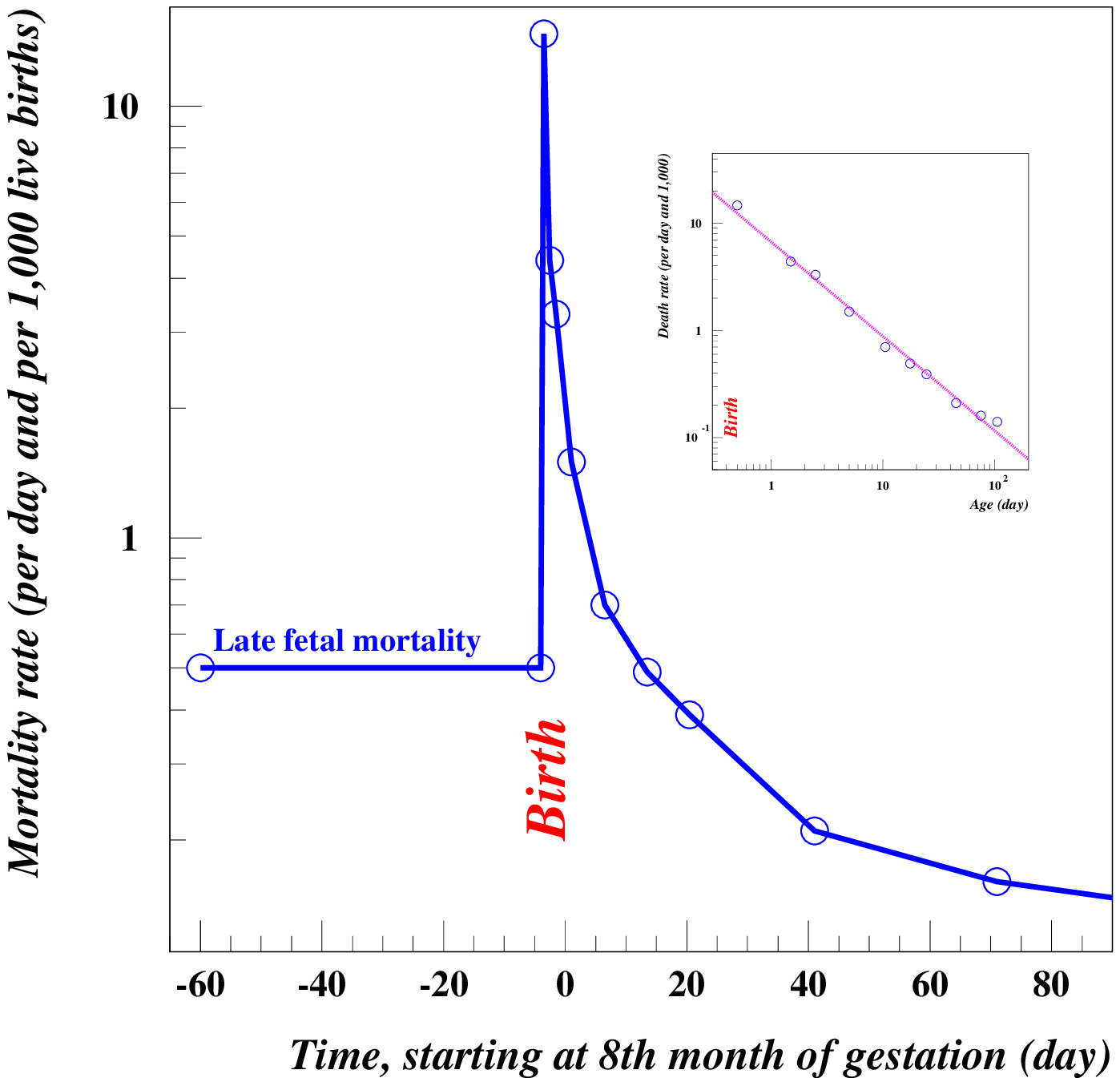}}
\qleg{Fig.\qhu 4\qhv Response function following 
the transition from gestation to birth. USA, 1923.}
{The graph shows the change in death rate which occurs 
in the wake of the birth transition.
During gestation the fetal death rate is fairly constant. Then,
following birth, ``defects'' which were not of great consequence during
gestation suddenly lead to a dramatic increase
of the failure rate. In terms of annual death rate, the peak rate
is about 3,500 times higher than the mortality rate in the
age interval 5-14. 
In the weeks following birth the death rate
has a power law decrease. For the inset log-log plot of the same data
the coefficient of linear correlation is $ 0.996 $
and the slope is 0.88.
It can be noted that there is a similar ``infant mortality''
pattern for other changes in living conditions and also
for the failures of technical devices following operation start.
As an example one can mention
electronic devices called VLSI (Very Large Scale
Integrated) chips (Wilkins 2002).}
{Source: Linder and Grove (1947 p. 574-575)}
\end{figure}

\qA{Suicide spike following imprisonment}

The fact of being arrested and incarcerated certainly marks
a dramatic change in living conditions, all the more so
when it happens to somebody for the first time.
It turns out that there is
a huge suicide spike in the first hours of incarceration.
Once converted into an annual rate, the rate is
about 100 times higher than in the general population. 
In subsequent days, a relaxation process sets in which slowly brings
down the rate toward the suicide rate in the general population.
%
\begin{figure}[htb]
\centerline{\psfig{width=12cm,figure=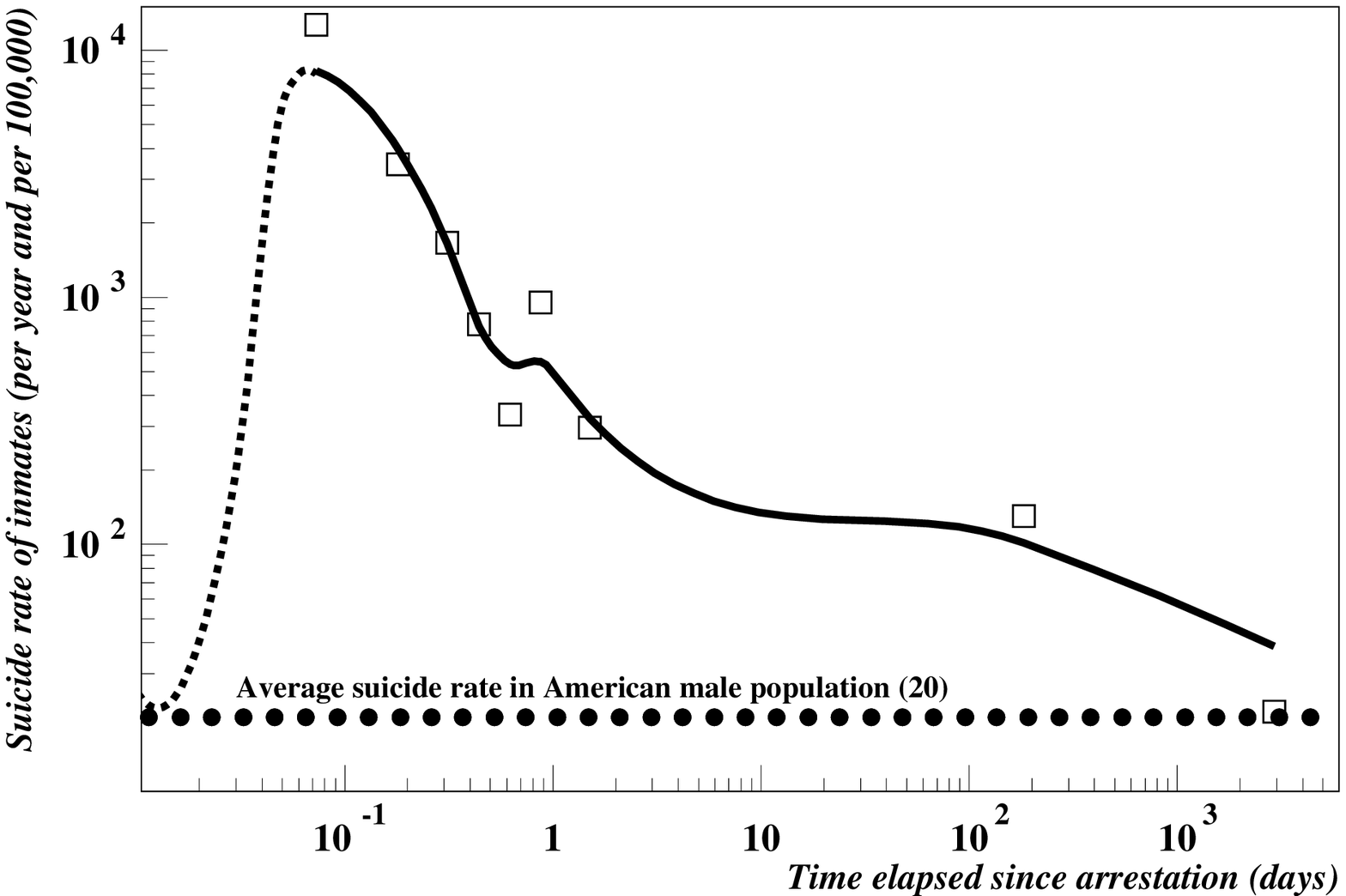}}
\qleg{Fig.\qhu 5\qhv Suicide rate of inmates as a function of the time
elapsed since their arrest.}
{The data are for the United States in the 1980s.
In the two days following incarceration, the suicide rate decreases rapidly.
Subsequently, the decrease continues at a much slower rate. After two years
the suicide rate becomes almost identical to the rate in the general male 
population. 
If one excludes the last two points the 
decrease roughly follows a power law:
$ y\sim 1/t^{\alpha} $ with an exponent $ \alpha=1.2\pm 0.4 $.}
{Sources: Hayes et al. (1988, p. 36); Roehner (2005 a, p. 669-670).}
\end{figure}
%

\qA{Suicide spike in the weeks following release from prison}

From the perspective of our transition conjecture,
there is a phenomenon which is even more revealing because less
expected. It is the fact that there is also a suicide spike in the
weeks after prisoners are released. As shown by the table below,
in the first week, the suicide
rate is some 8 times higher than 3 months after release.

\begin{table}[htb]

 \small 

\centerline{\bf Table 1\quad Number of deaths per week of ex-prisoners
in the weeks after their release.}

\vskip 3mm
\hrule
\vskip 0.7mm
\hrule
\vskip 2mm

$$ \matrix{
\tvi 
\hbox{Situation}  \hfill & \hbox{Suicide:}  \hfill & \hbox{Natural
  causes:} \hfill &
\hbox{Accidents:} \hfill \cr
\hbox{}  \hfill & \hbox{number of deaths}  \hfill & \hbox{number of
  deaths} \hfill &
\hbox{number of deaths} \hfill \cr
\qtb
\hbox{}  \hfill & \hbox{per week}  \hfill & \hbox{per week} \hfill &
\hbox{per week} \hfill\cr
\noalign{\hrule}
\qth 
\hbox{Week 1}  \hfill & 4.00\quad [1.00] &  4.00\quad [1.00] &
13.\quad [1.00] \cr
\hbox{Weeks 2,3,4}  \hfill & 1.70\quad [0.42] &  2.33\quad [0.57] &
8.3\quad [0.64]\cr
\hbox{Weeks 5-12}  \hfill & 0.87\quad [0.22] &  1.63\quad [0.40] &
6.2\quad [0.47]\cr
\qtb
\hbox{Weeks 13-24}  \hfill & 0.54\quad [0.13] &  0.66\quad [0.16] &
5.9\quad [0.45]\cr
\noalign{\hrule}
} $$
\vskip 1.5mm
Notes: The data are for the UK in the 1990s.
The numbers within brackets show the data in normalized form 
(first week=1). 
The decrease
in the number of suicides per week reflects a transient state
marked by a reorganization of social ties. 
For death by natural causes a factor which may play a role
is the fact that terminally ill prisoners are often released 
so that they can die in hospital (incidentally, 
this ``improves'' the death record of the prison).
\qL
Source: Sattar (2001, p. 34)
\vskip 2mm

\hrule
\vskip 0.5mm
\hrule

\normalsize
\end{table}

\qA{Transition from home to nursing home}

Numerous observations have shown that when people are moved from 
their home to an hospital or a nursing home, they experience a 
substantial increase in their death rate. Two cases will be reported
shortly which show that within a few months after admission
the increase can be as high as a multiplication
by 5 or 10. 
\qpar
Camargo et al. (1945) studied all first admissions of patients
over 65 years old to the state mental hospitals in Maryland.
Their average age was 74. 85\% of the patients were diagnosed as
psychotics. As reported in the graph below they found very high
death rates in the month following admission. 
\qpar

A similar study was performed in France by Th\'er\`ese Locoh (1972).
The average age at admission was 78 years old. The fact that these
persons were not psychotic patients certainly explains that the
death rates were about 5 times lower than in the previous study.
Nevertheless, the peak death rate is about 5 times higher than
the annual rate in the general population of same age.

\begin{figure}[htb]
\centerline{\psfig{width=15cm,figure=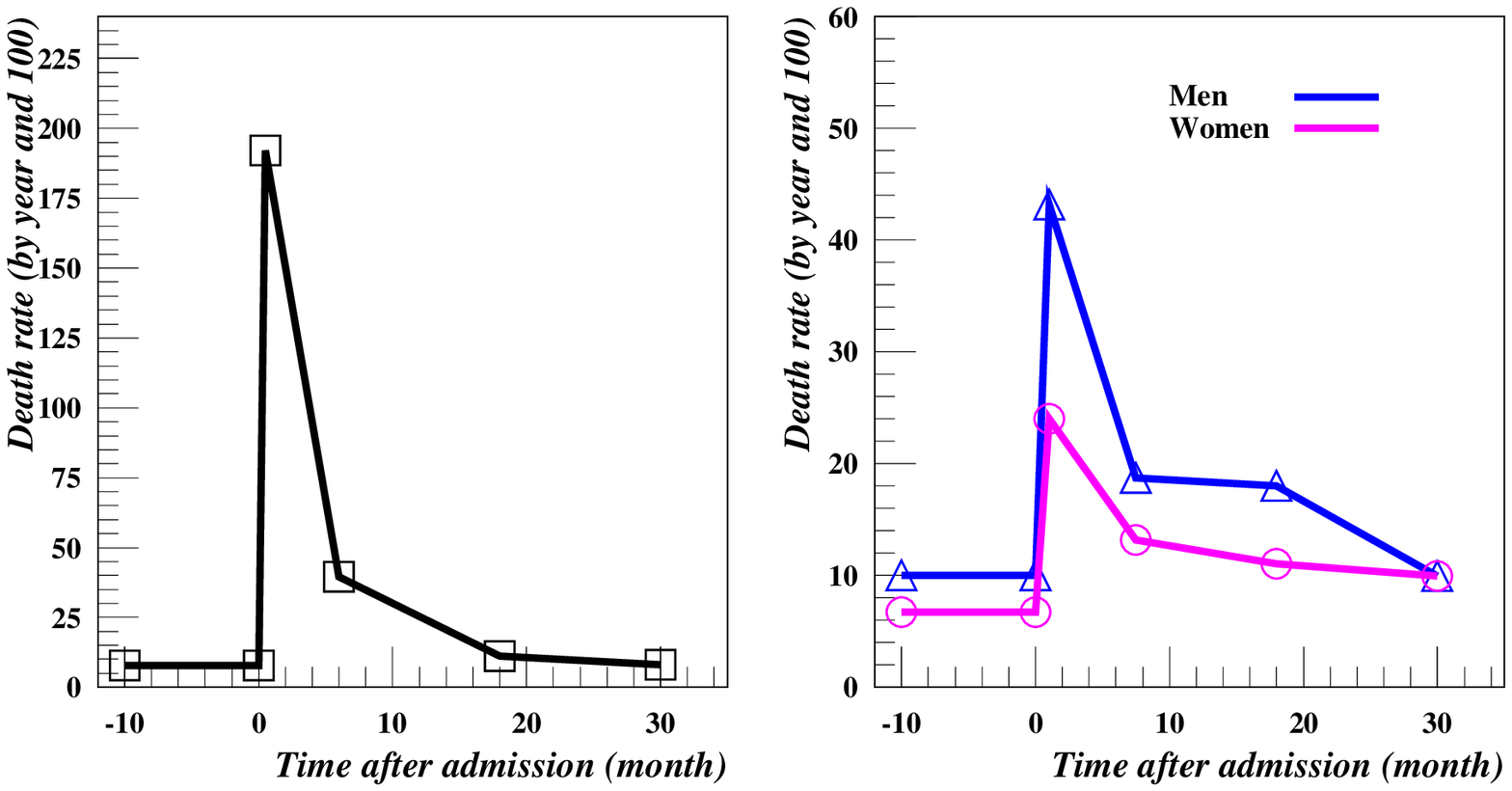}}
\qleg{Fig.\qhu 6a,b\qhv Mortality spikes in the transition from home 
to nursing home.}
{(a) Left-hand side: the data concern persons admitted to
state mental hospitals in Maryland; their average age was 73.9;
there were approximately as many males as females. The
very high rate reached in the first month is not a mistake:
16\% of the patients died in the first month and 
$ 12\times 16 =192 $. In the first year the mortality was about
15 times higher than in the general population of same age.
(b) Right-hand side: the data concern French persons admitted 
to nursing homes
in Paris and the vicinity; their average age was 78; there were
125 men and 480 women. The horizontal lines prior 
to time $ t=0 $
corresponds to the death rate of persons of same age in the general
population.}
{Sources: Maryland: Camargo and Preston (1945). Paris: Locoh (1972)}
\end{figure}

An objection has been raised by some authors.
They said: ``It may be that the transfer 
decision taken by the caregivers
(i.e. the relatives who are taking care of the
elderly persons) was motivated by a sudden deterioration of the
health of the persons''. If that would be the case the mortality
spikes shortly after admission would lose their significance in
relation with the transfer.
This objection can be answered in two different ways.
\qbu First, one should recall that it is within one or two months
after admission that the death spike is the most serious.
Such a time interval is very short
compared to the survival time of most of the diseases
which affect elderly persons (e.g. Alzheimer disease, cancer)
which is rather of the order of a few years. In other words,
even if the health of the persons had been declining it should not lead
to death so quickly. In addition, one should keep in mind that
most often the admission into a nursing home is 
subject to a delay
due to the existence of a waiting list. 
\qbu One must recognize that the previous argument is purely
qualitative and for that reason is not completely convincing.
A better answer is to focus on cases for which the transfer decision
is taken independently of the situation of the persons. That is
for instance the case when an institution is closed and all
patients must be transferred to other places. 
Two cases of relocation have been studied and reported in the
literature: Aldrich et al. (1963) and Killian (1970).
Mortality spikes after relocation were reported in both papers.
More details are given in Appendix B.

\qA{Transient mortality conjecture}

The previous observation leads us to the following statement.
\qdec{\it ``Any abrupt change in living
conditions generates a mortality spike which acts as a 
kind of selection process.''} 
This statement which will be called the {\it Transient Shock} conjecture
provides a qualitative model which leads to testable predictions.

\qA{Predictions}

The previous observations put us in the same situation
as a person who has seen the Sun rise around 7:00 am 
during 7 days. Naturally, he (or she) will also expect
a Sunrise around 7am in the following days. In order to make
this prediction the person does not need a mathematical
model of the solar system. Incidentally, such a model
would need a lot of inputs based on astronomical observations
(e.g. the minor and major axis of the orbits)
but nevertheless would not be able to predict the length of the day
which is a purely empirical parameter.
\qpar

Apart from the cases already tested, can we make 
other predictions? Here are a few examples.
\qbu For a person to lose his (or her) job is certainly a
major change in living conditions. Therefore, one would expect
a transient mortality spike in the weeks following the loss
of the job. Does it exist? The short answer is ``So far
we do not know''. 
There have been many studies about the suicide rates
of unemployed persons. For instance, Weyerer et al. (1995)
have conducted a long-term study. Unfortunately the
correlation between unemployment and suicide rates are
very low and hardly significant (under 0.20). 
In contrast one  would expect a 
significant suicide spike not in the 
long run but only in the weeks following the 
loss of the job. The problem is to find appropriate
statistical data.
\qbu For elderly persons, the fact
of  moving to another area far away from
their circle of relatives and friends is a difficult transition
for which one would expect a mortality spike.
\qbu For immigrants the fact of leaving their family behind
and moving to a country where people speak a different language
certainly represents a difficult transition.
Some statistical data are available for immigrants
who came to the United States at the end of the 19th century
or beginning of the 20th century (Nagle 1982, Mortality
Statistics 1910, pp. 586-595). It appears that 
for a broad sample of countries the suicide rate
of immigrants is much higher than either the rate in their country of
origin or the rate in the United States 
(see Appendix A and Roehner 2007, pp. 217-220).
Unfortunately, the data do not allow us to follow this phenomenon
in the course of time.
\qbu The transition single-to-married is a major change.
Therefore one would expect a mortality spike in the months
following the marriage. This is the purpose of the next section.
\qbu The transition married-to-widowed represents a major change.
Therefore one would expect a mortality spike in the months
following the death of the spouse. This point will be briefly
discussed at the end of the next section.

So far, we have mostly focused on death rates. In the
following sections we describe cases in which
one would expect transient marriage rates.
For instance, should we expect
a transient remarriage spike after widowhood?
We do not know yet.
When a widower wishes to get re-married he must first develop a 
network of friends. This may take some time. 
Up to now, the only information that we have is the long-term
remarriage rate of widowed persons. In 1960 in the United states
it was of same order as the
marriage rate of single persons (Grove and Hetzel 1968, p. 103).
However, around 1850 in the Netherlands, Switzerland and France
the remarriage rate of widowers was 3 to 4 times higher
than the marriage rate of single persons (Bertillon 1879).

\qI{Mortality spike in the transition ``single-to-married''?}

\qA{Why is the question of importance?}

\begin{figure}[htb]
\centerline{\psfig{width=8cm,figure=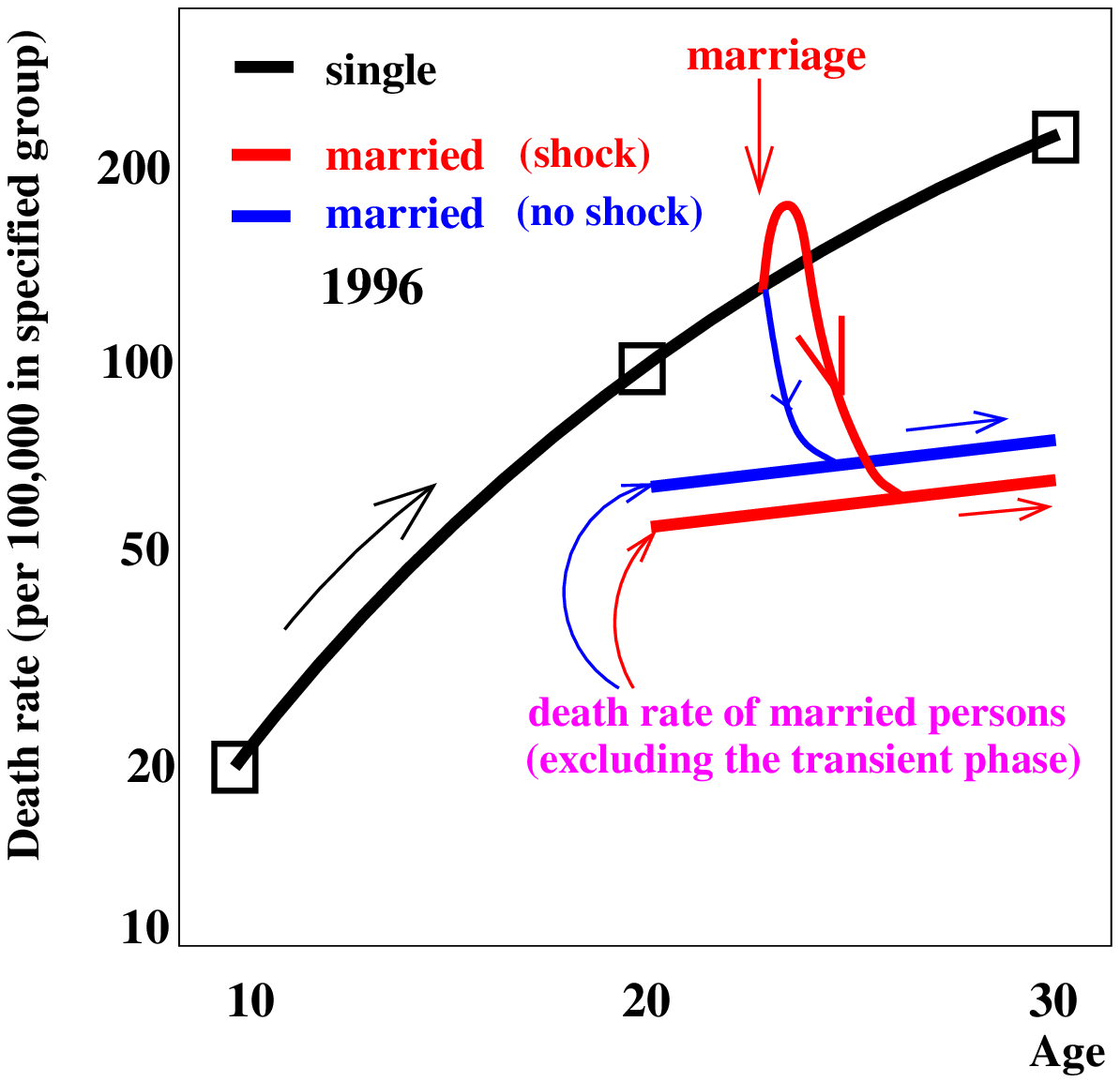}}
\qleg{Fig.\qhu 7\qhv Is there a mortality spike in the aftermath of
marriage?}
{The graph depicts the death rate change in the single-to-married
transition. There are two possibilities.
(i) Smooth transition (in blue) (ii) Transient shock (in red)
during which
the death rate climbs above the rate of single persons.
Incidentally, it can be observed that the
increase with age of the death rate of single persons
is much faster than the corresponding increase for married
persons, a point which was already highlighted in Paper 1.
The data used to draw the figure are for 1996 but
the picture would be similar for any other year.}
{Source: National Vital Statistics Report, Vol. 47, No 9, 10 Nov 1998,
Table 21, p. 73.}
\end{figure}

In everybody's life
marriage is certainly a major transition. We know that its
long-term effect is to reduce mortality rates by a factor 2 or 3.
We attributed this effect to increased interaction. 
However, this does not necessarily exclude a transient mortality
spike in the months following the marriage because it
may be that some persons who were well adapted to their non-married
status will not be adapted quite as well to their new marital
status. The question of
whether or not there is a transient shock following
marriage is an important point because if the answer is ``yes'' it
will show that mortality shocks do not only occur as a result of
bereavement but in fact are brought about by any major change
in living conditions.
\qpar

\qA{Design of the observation}

How can we find out whether or not such an effect exists?\qL
It is not an easy task because we are looking for a short term effect;
typically, one would expect a time constant of a few months.
Moreover, even if the spike reaches a level twice as high
as the death rate $ d_s $ of single persons, it will remain low because 
at that age $ d_s $ itself is fairly small. This means that to
get an insight we need large samples.
\qpar

As always, there are two methods, namely longitudinal or transversal
analysis%
\qfoot{Transversal (also called cross-sectional) studies refer
to the analysis of data collected at a specific point in time.}
.
\qbu In principle,
longitudinal analysis is possible only in countries where 
a cohort can be followed in the course of time through the
civil registry statistical system, as is for instance the case
in the Scandinavian countries. It can be observed that
in such a framework 
the present inquiry should be considerably
easier than for the deaths of widowers because 
in any month the number of 
newly married people is about 1,000 times larger than the number
of newly widowed persons of same age. 
\qbu For transversal analysis one needs to design a suitable 
``experiment''.
A possible methodology based on
the monthly distribution of deaths will be tried below,

\qA{Semi-longitudinal analysis}

Usually census data cannot be used to do longitudinal
analysis because they provide a picture of the population
at a given moment. Few questions are usually asked
about the past. There is a good reason for that
which is the fact that the recollection of people about
events that are more than 2 or 3 years old is not
very reliable.
\qpar

Here we will make use of a census question%
\qfoot{According to the website of IPUMS, the variable
MARRINYR identifies persons who had married within the 12
months preceding the date of the ACS (American Community
Survey) interview. Although this survey is carried out monthly.
the Census Bureau
does not make publicly available the interview months of 
respondents. Protection of confidentiality is the reason
that is put forward despite the fact that it is not obvious why
knowing the month of the interview should be an issue. 
It can be observed that, although the term 
``interview'' is routinely used,
in fact, the questionnaires are sent out by post or Internet
(real interviews are conducted only in a number of special
cases). This suggests another reason for
not giving the month: although the questionnaires 
are sent out on the same day,
the dates of the replies may cover a fairly broad range,
a range possibly extending over more than one month.}
that concerned the 
{\it recent past}: ``Did you get married within the last year''.
If we select the persons who responded ``yes'' and if among them
we select the persons who declared to be widowed at the 
time of the census interview, we will get the persons
who got married and then lost their spouse within 12 months. 
In other words, we will be able to compare the death rate
of married persons in their first year of marriage to
the long-term death rate of married people.
\qpar
The results of this observations are summarized in the graph below.
The question about marriage within the past year was asked
only in a few censuses, basically the census of 1880 and
the ``American Community Survey'' after 2006. This
survey is conducted every month on a different sample of 
about 300,000 persons. 
Over the whole year, it concerns about 3 million
persons.  
%
\begin{figure}[htb]
\centerline{\psfig{width=16cm,figure=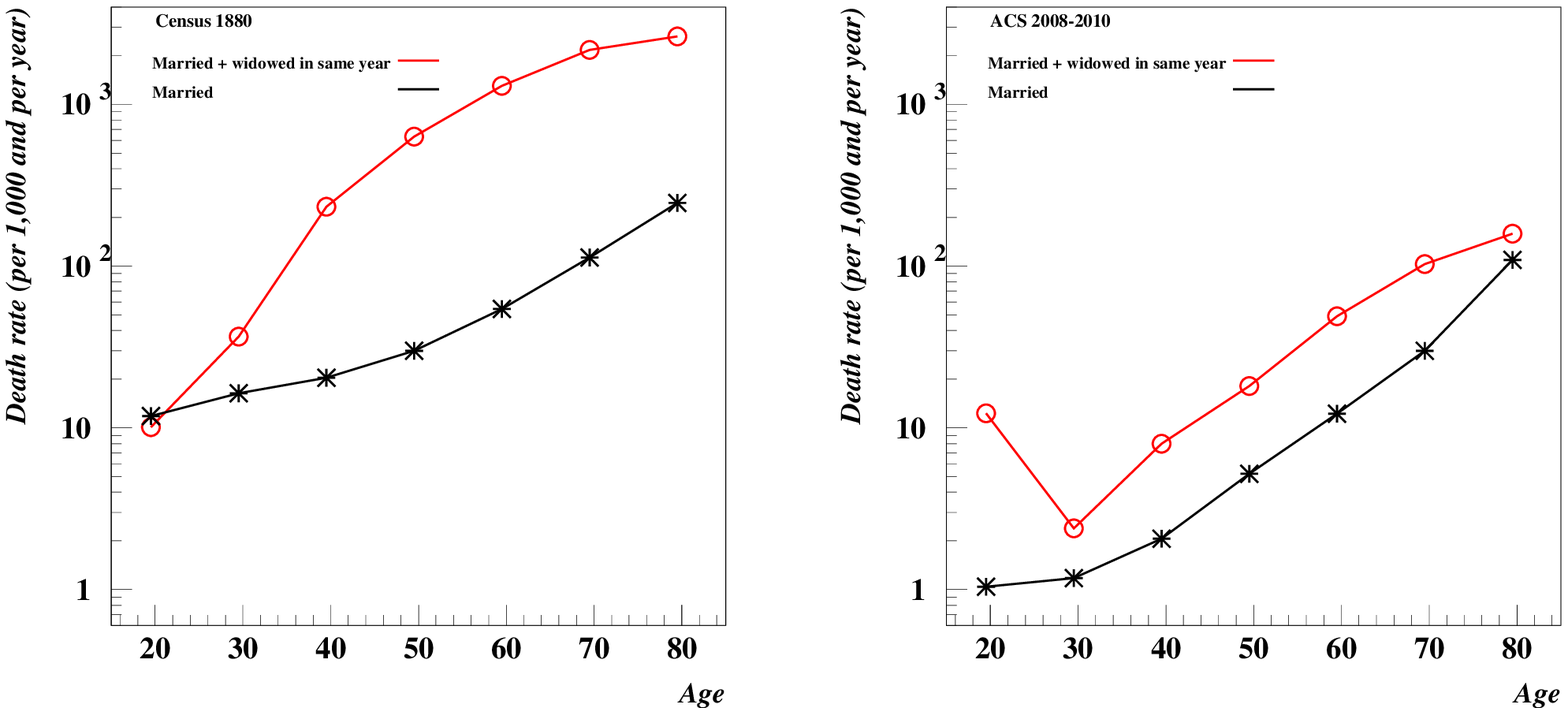}}
\qleg{Fig.\qhu 8\qhv Death rate in the first year of marriage.}
{As most marriages last several decades the overall rate
for married people (in black with stars)
is a long-term rate as opposed to the other
curves (in red) which are a short-term rates over the
first year of marriage.
Needless to say, the persons who got married and became
widowed in the same year are in fairly small number.
In the 5\% sample of the 
census of 1880 they were 579 and in the ACS 2008-2010
survey they were 397. It can be seen that the death rate
in the first year of marriage is consistently higher
than the overall death rate of married persons.}
{Source: ACS 2010-3 years (2008-2010) extracted from the IPUMS
database. The file comprises 9,093,098 lines (each line
corresponds to a different person).}
\end{figure}
%
The definition of the death rate of the persons married
and widowed in the same year requires some explanations.
\qpar
For a given age interval, 
among the $ n_1 $ persons who got married in the past year we
select the $ n_2 $ persons who became widowed. 
The ratio $ r=n_2/n_1 $
represents a kind of death rate; however its real definition must
be considered more closely.
Firstly, it is obvious that the death of the partner occurs
within a time interval which is smaller than one year. 
If the marriage occurs toward the end of the year, it must
be followed almost immediately by the death of the partner.
In order to define this death rate more completely we need
to know what is the average time interval between marriage ($ M $)
and widowhood ($ W $). In the absence of any contrary evidence 
it is natural to
assume that over 365 days, both $ M $ and $ W $ are distributed
uniformly. However, $ M $ and $ W $ are not independent random
variables, because $ M<W $. A simulation reveals that in such a 
situation the average $ E(W-M) $
of the time interval between $ W $ and
$ M $ is one quarter of the  year. Thus, in order to
express $ r $ as an {\it annual} death rate, 
it must be multiplied by a factor $ k_1=4 $.
\qpar

There is a second correction which needs to be done.
The ratio $ r $ is not really a death rate per individual, but
rather a rate for the end of marriage. As the death of {\it any} of
the two partners will result in ending the marriage, we see
that $ r $ actually refers to twice the death rate per person.
In order to express $ r $ as a rate per individual it must
be multiplied by a factor $ k_2=1/2 $. Thus, altogether,
$ r $ must be multiplied by $ k_1k_2=2 $. 
\qpar

On the graph for 1880 we can see that the death rate almost
reaches a level of 1,000 per thousand. This should not be 
surprising. Rates higher than 1,000 simply mean that the whole group
died in less than one year. For old age
the observed values of $ n_1,n_2 $ indeed reveal a high mortality.
For instance, in the last age group, namely $ 75-84 $, of the
$ 93 $ persons who got married in the past year, $ 61 $ became widowed
in the interval between their marriage and the census interview.
By contrast, in the 2008-2010 survey
the $ n_2/n_1 $ ratios are much lower. Thus
in the last age-group, there were $ 782 $ persons who got
married in the past year and only $ 31 $ became widowed prior
to the interview.

\qA{Longitudinal analysis}

Usually, censuses do not give any information about death events
for the simple reason that the persons who are able to respond are
still alive. However, in a few cases a question 
was asked about possible death events of relatives. That was the case
in the American Community surveys which were made after 2006.
The question was ``Have you been widowed in the past year?''.
Through another question we are also able to learn for how
many years the persons had been married. Taken together, these
data allow us to follow the persons in the course of time and
to perform a longitudinal analysis.  
\qpar
We considered 5 cohorts $ C_i $ of married persons who got 
married in 2000, 2001, 2008, 2009 and 2010 respectively. Their death rate
was computed as $ d_i=D_i/C_i $ where the
$ D_i $ denote the number of spouses who became widowed
in 2010.
The $ D_i $ are of the order of a few dozens for young ages but
become of the order of only one dozen for old ages. In order
to reduce the statistical fluctuations which come along  with such
fairly small numbers, we lumped together the data for the
cohorts $ 2000-2001 $ and $ 2008-2009 $. The results are shown in 
the graph below. One sees that at a given age the death rate decreases 
with the length of time spent in married status,

\begin{figure}[htb]
\centerline{\psfig{width=16cm,figure=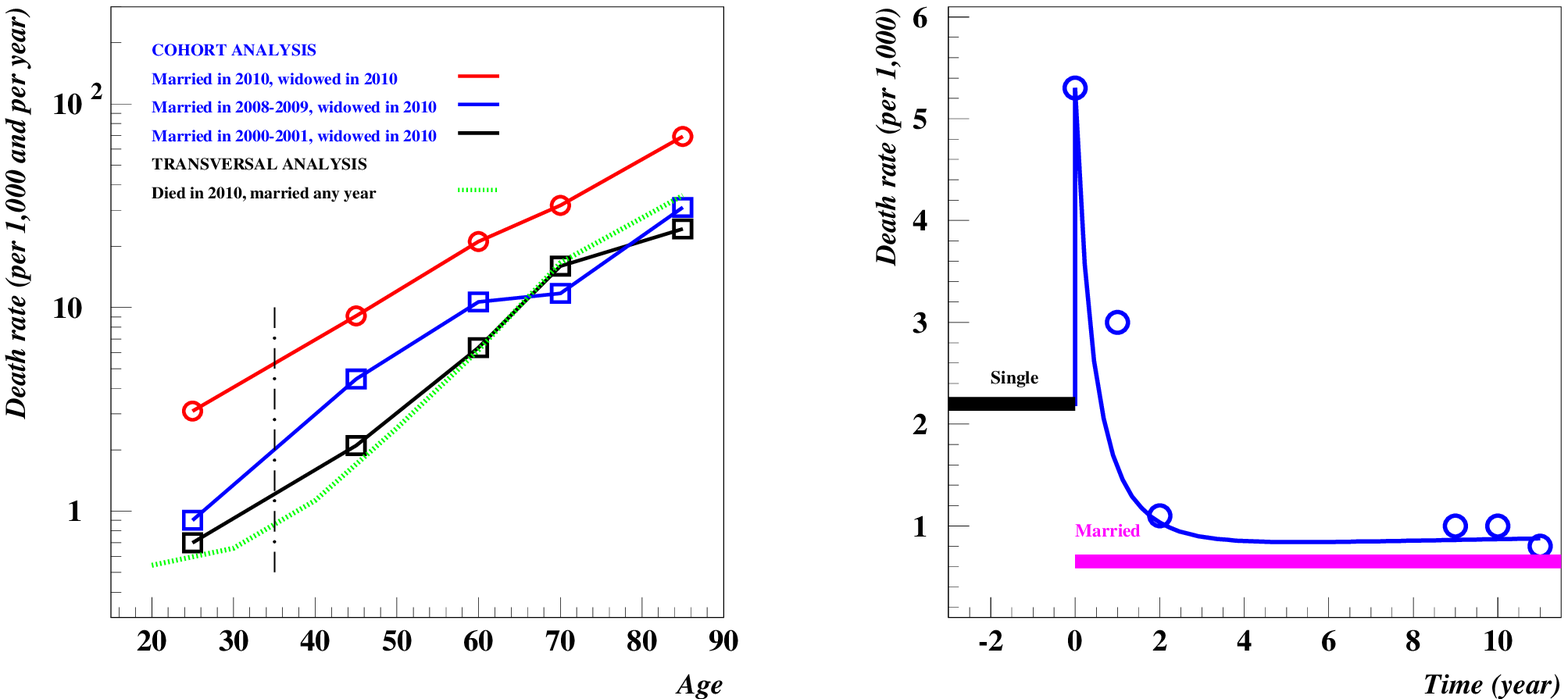}}
\qleg{Fig.\qhu 9a,b\qhv Death rate for married people as a function
of age and length of marriage.}
{(a) The green dotted line gives the long-term death rate of
married persons. It almost coincides with the death rate of 
people who have been married for 10 years (black line).
On the contrary,
the cohort of the persons who got married
and widowed in the the same year (red line)
displays an inflated death rate,
especially for young ages. The vertical dashed line at age 35 
corresponds to the section shown
in the figure on the right-hand side. (b) The figure shows
the death rates experienced by a cohort of age 35
that gets married at 
time $ t=0 $. Immediately after marriage, there is a 
death-rate spike whose time constant is of the order of one year.
Then the death rate converges toward its steady state average
which (as we know from Paper 1)
is lower than the death rate of never-married persons.}
{Source: The cohort analysis is based on the AC survey of 2010.
The data are available on the IPUMS (Integrated Public 
Use Microdata Series) website of the 
Minnesota Population Center, University of Minnesota.
This ACS-2010 (1 year) file comprised 3,061,735 persons.
The transversal analysis which lead to the curve with the stars
is based on data given in the
``National Vital Statistics Report'' of 8 May 2013 (p. 8).}
\end{figure}
\qpar
As the existence of a mortality spike after marriage is
rather counterintuitive, we need to discuss the reliability
of census data.
Censuses are conducted with great care but ultimately their
accuracy depends on whether or not the respondents answer
the questions correctly. In the present case, 
what is to be expected regarding the ratios $ d_i=D_i/C_i $?
One expects that $ D_i $, the number of persons who became
widowed in the year of the survey, will be fairly accurately
reported because these are recent events in the respondents' lives.
The same observation holds with respect to marriage
for the persons who got married
in 2009 or 2010. For more distant years such as 2000 and 2001
the recollection is probably not so good. However, for those years
the data can be checked through a comparison with transversal
data. The comparison shows that the two sets of data are fairly
consistent with one another.

\qA{Transversal analysis}

\begin{figure}[htb]
\centerline{\psfig{width=16cm,figure=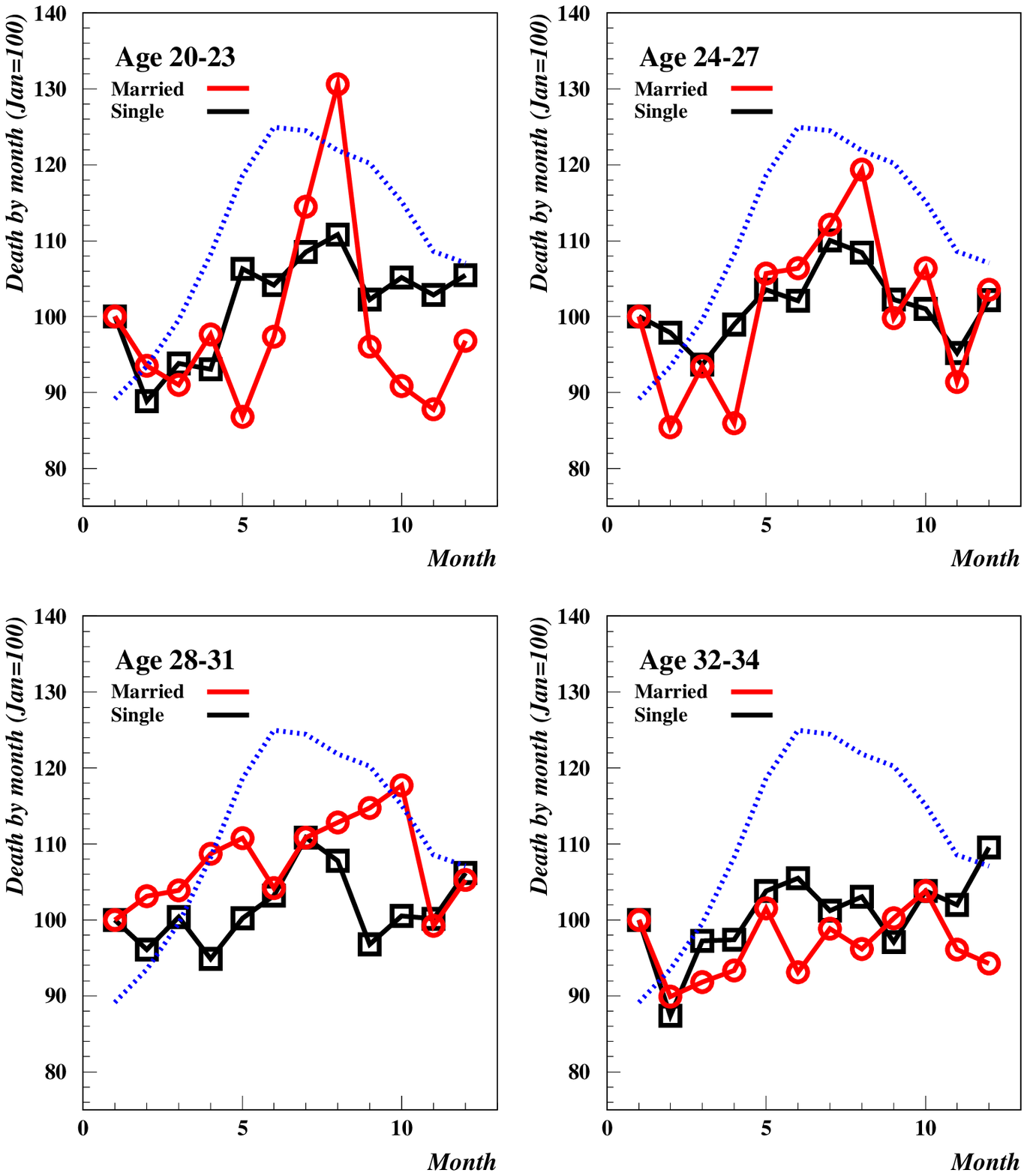}}
\qleg{Fig.\qhu 10\qhv Is there a transient mortality spike in the
months following marriage?}
{Horizontal scale: month (1992). Vertical scale: Number
of deaths by month for single (black line with squares)
and married persons (red line with circles) respectively.
The panels show successive age groups. In each age groups there
are several hundreds deaths (the smallest number, namely 182, is in
the 20-23 age group of married persons).
The (blue) dotted curve shows the monthly number of marriages.
It can be added that
for older age groups from 35 to 44 (not shown) 
the death numbers of married people do not display any peak.}
{Source: Mortality detail file for 1992 issued by the CPSR
(Consortium for Political and Social Research).
The file comprises all deaths that  occurred in the US in 1992
(altogether there were 2,179,187 deaths).}
\end{figure}
From the two previous subsections
we know that there is a mortality peak in the 
first year of marriage. The TS conjecture would suggest an
even taller spike in the one or two months immediately
following marriage. To explore this further is our objective in this subsection.
\qpar
The idea on which the present transversal analysis is based is
to use the seasonality pattern of the marriages.
With a uniform distribution of monthly marriages the prospect
would be hopeless because it would be impossible to distinguish
the deaths due to the transient effect from all others. 
Fortunately,
in the United States the monthly
frequency of marriage peaks in May and June. 
The marriage rate in May-June is about twice the marriage rate
in January. These monthly fluctuations are much larger than 
the seasonal fluctuations of the
death rates for which the max/min ratio
is about 1.2 (with the maximum in winter and the minimum in summer).
Thus, if we see
an increase in death numbers in the months following
May-June, it may be attributed to a transient death effect
triggered by the marriages.
Of course, in order to be convincing such an increase must occur
in the age groups in which the marriage rate is highest. As
a further confirmation test it
should not be seen in age groups over 35.
\qpar

What data do we need for that investigation?
We need monthly mortality data by age and marital status.
Monthly mortality data are indeed available in the ``Vital
Statistics of the United States'' but they are not broken
down by age and marital status. Fortunately,
the ``Inter-university Consortium for Political and Social Research''
(ICPSR) provides a file which will solve our problem.
It lists {\it all} the deaths which occurred in 1992 and
for each of them it gives the age, marital status, cause of death
and many other characteristics. The downloaded 
ASCII zip file had a size of about 50 M; once inflated
the size of the file containing the data reached 333 M!
\qpar

In the summer months (i.e. in the 2 or 3 months following May-June)
there appears to be a transient excess mortality of
married persons. Moreover this excess mortality does not exists for ages over
35. \qL
It must be kept in mind that for a fraction of 
the death certificates
of married persons who die shortly after their marriage the new
marital situation may not have been updated. The death registration
procedures may also be different from state to state, but one should 
not be too surprised to see a May-June
peak in the deaths of single persons

\qA{Comparison of the three methodologies}

Each of the three methods that we used has 
its merits and its shortcomings.
\qbu The first set of data had the advantage of existing
for a census of the 19th century. This gave us the opportunity
to check whether the effect exists over a broad time range.
\qbu The second set of data allowed us to probe different
lengths of time spent between marriage and death.
\qbu The results obtained through the third method 
were more ``noisy'' than those of the two other methods%
\qfoot{This is of course partly due to the fact that
in this approach we did not use a logarithmic scale. 
When it is the age which is the independent variable 
the death rate differences for successive age intervals
are so large that the random fluctuations become
invisible.}%
,
but this approach had the merit of suggesting 
that there may indeed be a 
tall and sharp death rate peak within one or two months
after marriage.

\qA{Transient shock after widowhood}
 The death of a spouse results in a drastic change in living
 conditions. The ``Transient Shock'' conjecture would lead us
to expect 
a mortality spike in the months following widowhood. 
The transition from single to married status implies
(i) a change in individual living conditions (ii) a 
re-arrangement of ties: weakening of the ties with
parents and strengthening of the ties with the spouse.
Similarly the transition from married to widowed status
implies (i) a change in individual living conditions
(ii) a severance of the bond with the spouse and possibly 
a strengthening of the links with relatives or friends.
In addition there is the psychological dimension of
the bereavement process. 
\qpar

Although a more ``plausible''  effect than the marriage shock,
the widowhood shock is much more difficult to study statistically.
The effect 
is difficult to measure for the simple reason 
that the deaths of young widowed persons are much rarer than the 
marriages of young persons. Of course, the death of elderly
widowed persons are not rare but in that case it is difficult
to separate the mortality due to aging from the mortality
due the widowhood effect. In other words,
for elderly widowers the effect
is blurred by a high level of ``background noise''.
\qpar

Many studies have attempted to analyze the widowhood effect.
Several of them are summarized in Table 2.

%
\begin{table}[htb]
\centerline{\bf Table 2 \quad Studies of the widowhood effect
  according to the number of deaths of widowed persons}

\vskip 3mm
\hrule
\vskip 0.7mm
\hrule
\vskip 1.2mm

\color{black} 
\small

$$ \matrix{
\hbox{Reference} \hfill &\hbox{Year} &\hbox{Age} & \hfill  \hbox{Deaths of}\cr
\qtb
\hbox{(first author)} \hfill &\hbox{of paper}&\hbox{} 
& \hfill \hbox{widowed persons} \cr
\noalign{\hrule}
\qth
\hbox{Mendes De Leon } \hfill & 1993   & >65& \hfill 22 \cr
\hbox{Bojanovsky (*)} \hfill & 1979,1980 &\hbox{any age} & \hfill 181 \cr
\hbox{Young } \hfill & 1963  & >55& \hfill 906 \cr
\hbox{Schaeffer} \hfill & 1995 & >40 & \hfill 934 \cr
\hbox{Kaprio} \hfill & 1987  &\hbox{any age}& \hfill 7,635 \cr
\hbox{Martikainen} \hfill &  1996& >40 & \hfill 9,935 \cr
\hbox{Mellstr\"om} \hfill & 1982  &\hbox{any age}& \hfill  360,000\cr
\qtb
\hbox{Thierry} \hfill & 1999 & >35 & \hfill 819,000  \cr
\noalign{\hrule}
}
$$
\vskip 2mm
Notes: (*) This study differs from the others in the sense that
it considered only death by suicide. 
{\it }
\vskip 2mm
\hrule
\vskip 0.7mm
\hrule
\end{table}

Not surprisingly, the most successful study so far has been
the one by Thierry (1999). It provided clear evidence of a
mortality peak in the first year after the death of the 
spouse. Unfortunately, because the data that he has been
using give only the year of 
widowhood (and not the exact date) this study cannot tell
us what happens during the first months of widowhood.
We will leave this question open for a further investigation.

\qA{Conclusion}

The next step would be to build a mathematical model.
However, we believe that one should not hurry to do that without
a more thorough understanding. So far, we have
an overall grasp of
the transient shock effect but there are several points
which remain unclear. For example,
the relaxation mechanism, in almost all cases for which sufficiently
detailed data were available (birth, arrest, marriage) the death rate
seems to decrease as a power-law instead of an exponential.
We would like to confirm this feature by studying additional 
cases and as a second step understand the mechanism behind it.

\appendix

\qI{Appendix A. Effect of external factors on suicide rates}

 In the subsections entitled ``Suicide spike following inprisonment''
and ``Suicide spike in the weeks following release from prison''
we considered the effect on suicide rates of drastic changes in living
conditions. In the present appendix we wish to complement
these observations with two others: (i) effect of immigration
(ii) effect seasonal factors. 
While immigration certainly entails a transformation 
in living conditions, it is a change which is less severe than
for inprisonment. The change brought about by seasonal factors
is even more moderate. Nevertheless, it turns out
that these changes have
visible effects on suicide rates. 

\qA{Suicide after moving to another country}

To be taken into custody is only one of several possible
mechanisms
leading to a disruption of social ties; the process of emigration 
is another one. 
Back in the 19th century when an
individual or a family emigrated from an European country (say Italy
for instance)
to the United States it implied a sharp interruption in the contacts with
relatives and friends left behind. Furthermore, 
until the language barrier
was surmounted it was not easy to establish social links with American
people, except of course with other Italian immigrants. 
\qpar
%
\begin{figure}[htb]
\centerline{\psfig{width=15cm,figure=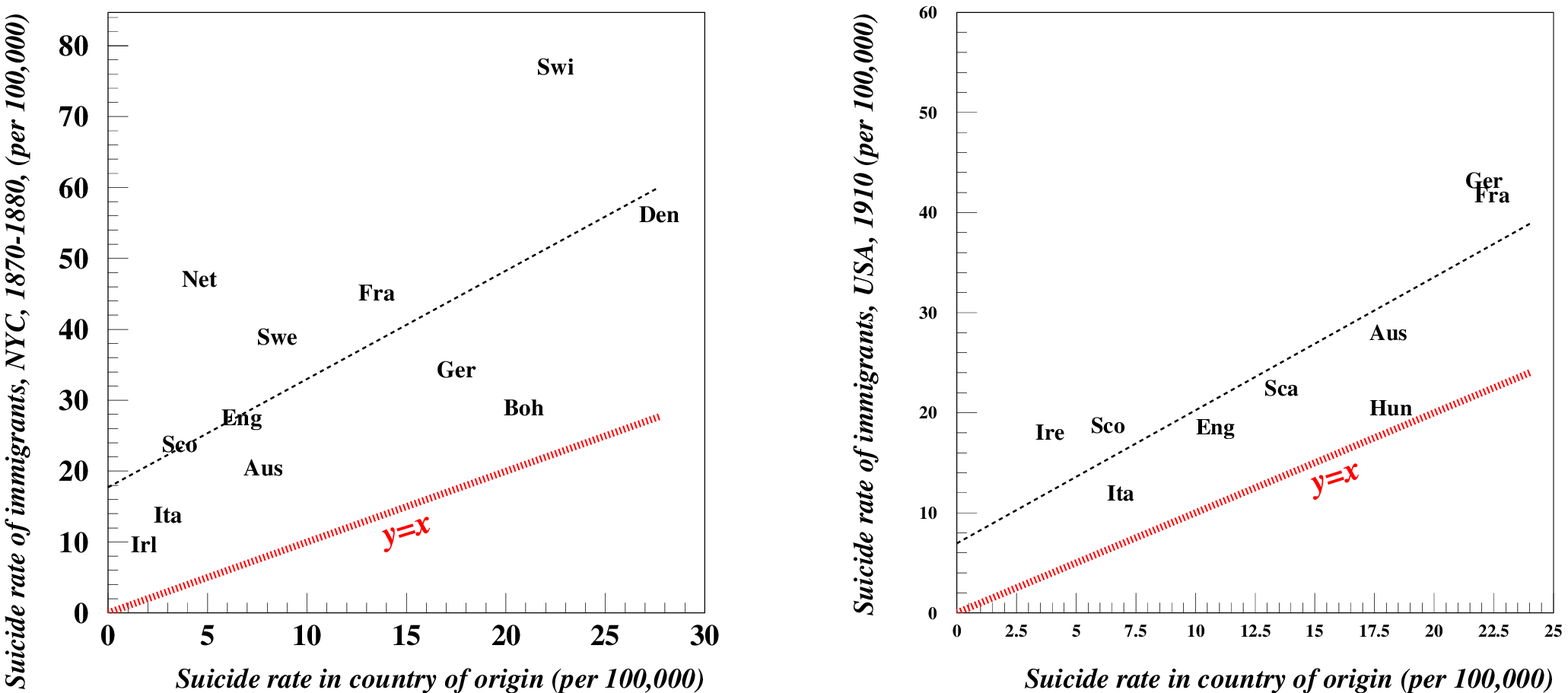}}
\qleg{Fig.\qhu A1a,b\qhv Suicide rate after relocation.}
{The two graphs compare the suicide rates of immigrants 
after they relocated into the the United States to
their suicide rates in their home country.
``Aus'' stands for Austria, ``Boh'' for Bohemia, 
``Hun'' for Hungary,
``Sca'' for Scandinavia, the other labels are self-explanatory.
The word ``immigrants'' refers to all persons born in a foreign
country.
The graph on the left-hand side is for New York City 
in the decade 1870-1880 (correlation is 0.70)
while the one on the right-hand side is for the United States in 1910
(correlation is 0.85).}
{Sources: (a). Suicide rates in NYC: Nagle (1882), suicide rates
in countries of origin: Durkheim (1897), Krose (1906).
(b) Suicides in the US.: Mortality Statistics (1910, pp. 586-595);
foreign born population by country of birth: Historical Statistics of the U.S.
(1975, p. 117); suicide rates in country of origin: 
World Health Organization (1996).}
\end{figure}

From a network perspective one would expect the suicide rate of immigrants
to be inflated by the weakening of their social ties. 
The graphs (Fig. A1 a,b) show that 
the data points are above the line $ y=x $ which means that
the suicide rates of immigrants are higher than in their respective 
countries of origin. In fact, they are also much higher than the
suicide rate in the general US population. This suggests the existence
of a shock effect although in the present case we do not know
its time constant.
\qpar

Two additional comments should be made.
\qbu The first comment is in answer to a possible objection.
It is well known that in European countries the suicide rates
are higher for men than for women. Often the ratio is about 2.
In the countries of origin there were of course as many males
as females. Unfortunately,
we do not know the exact sex ratios of immigrants
for each separate country; the only available information 
is the global sex ratio for all immigrants in a given year.
Thus, in 1875 there were 61\% males while in 1910 their
percentage was 73\% (Historical Statistics of the United States
1974 p. 112).
Can this difference in sex ratios explain
that the suicide rates are higher in the US? The answer is no.
Replacing 50\% by 61\% or 73\% would result in small adjustments
of the suicide rates:
some 17\%  in the case of a population composed of 73\% males.  
However the rates of immigrants  are {\it much
larger} than in the countries of origin: at least 50\% 
and many are of the order of 100\% (see for instance the cases
of England, France or Germany).
\qbu Apart from suicide,
the ``Mortality Statistics'' of 1910 also gives death data 
by country of origin for 28 other causes of death. A natural
question therefore is whether the previous analysis can be repeated
for other causes. Unfortunately, it would have little
significance for the following reasons. Infectious diseases
(e.g. typhoid fever, tuberculosis, pneumonia) are dependent
upon local conditions which means that higher rates in the US
may be explained by fairly poor living conditions of immigrants.
On the other
hand, the incidence of non-infectious diseases such as cancer
or heart disease is highest in old age. Yet, 
at their arrival in the US about 85\% of the
immigrants were in the age group 15-44. In other words, death
by cancer or heart disease will mostly affect persons who have
been in the US for 20 or 30 years and for whom the transition
shock is old history. 

\qA{Sensitivity to exogenous factors increases with age}

As is well known, suicide rates display a seasonal pattern,
with a minimum in December and a maximum in April-May.
In spite of the fact that this effect has been known for over
a century and a half, we do not yet know on what mechanism 
it relies. The only clear conclusion is 
that monthly suicide rates
are significantly correlated with the length of the day. 
For our present purpose the only assumption that we need to
make is that the seasonal pattern is due to exogenous
factors. 
Then, keeping in mind the conclusions drawn from
our former observations about the frailty of elderly persons
one would expect such persons to
be more sensitive to seasonal factors than are younger persons.
\qpar
In order to see if this is confirmed by observation, we computed
the correlation
$ C(\hbox{age}) $ between monthly suicide numbers in a given age interval
and the length of day. It turns out that for 
young persons, say under 30, $ C(<30) $ is almost zero.
On the contrary, $ C(>60) $
is quite significant which means that the suicides of persons over 60
display a clear seasonal pattern.
\qpar
These results are summarized in Fig. A2. It shows that
$ C(\hbox{age}) $  increases fairly steadily with age.

%
\begin{figure}[htb]
\centerline{\psfig{width=8cm,figure=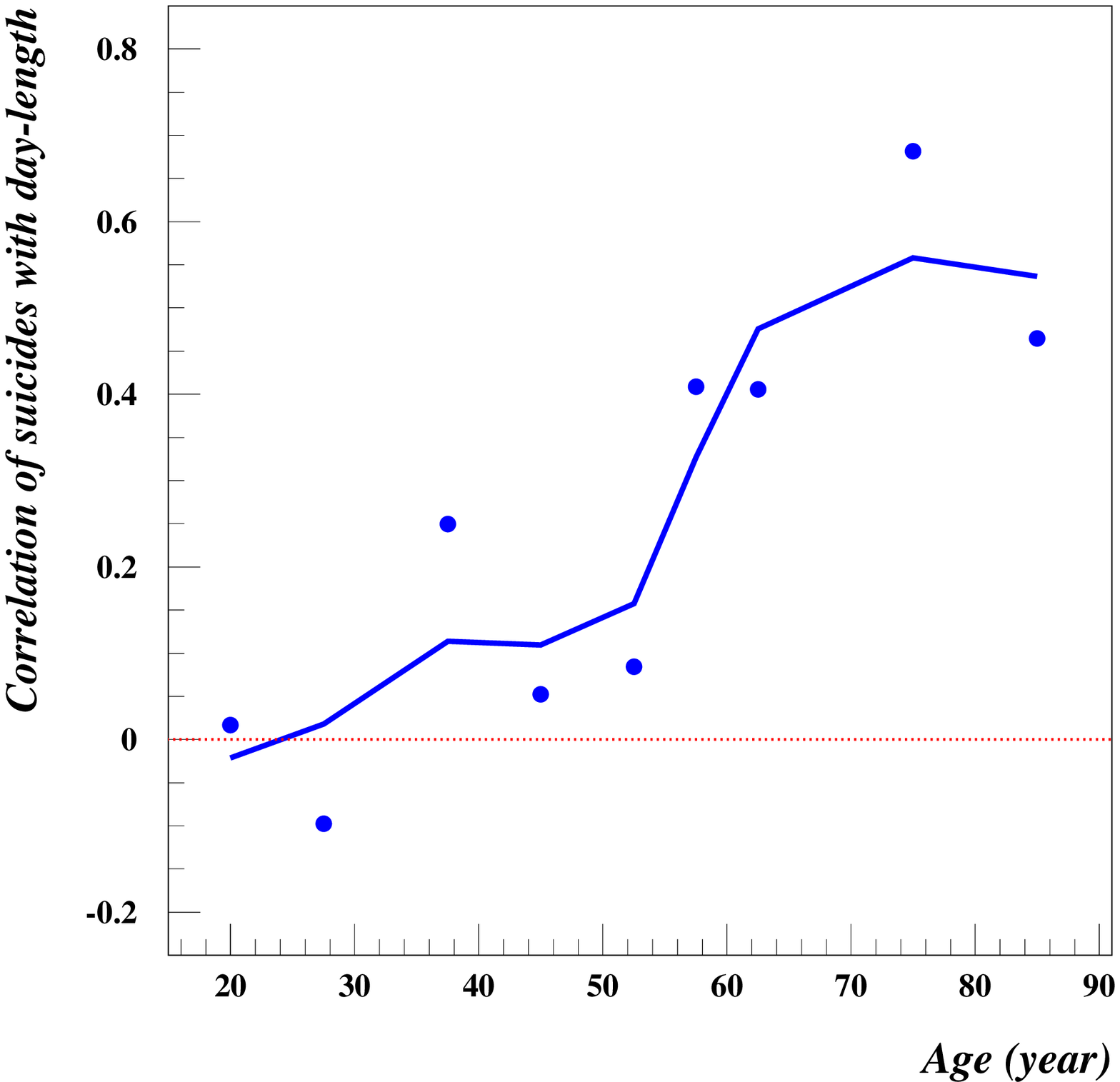}}
\qleg{Fig.\qhu A2\qhv Correlation of monthly suicides with day-length,
Germany, 2004}
{The correlation with day length is a measure of the amplitude
of the seasonal pattern of suicide rates. The graph shows that
whereas the seasonal pattern is almost nonexistent for young persons
its amplitude steadily increases with age. The same kind of evidence
is available for the United States.}
{Sources: Statistisches Bundesamt}
\end{figure}

If one wishes a physical analogy for this effect one 
can consider the following situation. 
Objects belonging to the Kuiper Belt which is located beyond Neptune
experience a weak solar attraction; as a result, their trajectories
will be strongly affected by the movements of their nearest neighbors.
On the contrary a planet like Mercury has a strong gravitational
bond with the Sun and will be little affected when a comet comes
close to it. 
\qpar
 
One must recognize that at this point the explanation 
regarding the seasonal effect
is rather a conjecture. Before it can be accepted, it must be
confirmed by supplementary evidence. For instance,
our previous argument would lead to the prediction
that elderly immigrants
experience a higher jump in suicide rates (with respect
to the rate in their country of origin)
than young immigrants. 
Unfortunately, this prediction is difficult to test.
The obstacle is not only that the required data may not be available,
rather they may not exist. Indeed, historically there were much less
elderly immigrants than young immigrants which means that the
numbers may just be too small to provide reliable suicide rate
estimates.

\qI{Appendix B. Former studies of elderly people's relocation}

In the subsection entitled ``Transition from home to nursing home''
we described the results obtained in two studies, namely Camargo
et al. (1945) and Locoh (1972). In the present appendix we wish
to review a number of other landmark studies. In particular
we will try to explain how, over the past three decades, 
the interest of US researchers shifted away from the investigation
of mortality effects and concentrated instead on more qualitative
features.
\qpar
First, we wish to describe in more details the studies
already mebtioned about the relocation of elderly persons.

\qA{Mortality spike in the relocation of elderly patients. 
Observation 2 (1968)}
In 1968 the Stockton hospital in California decided to close
part of its facilities for elderly persons. As a result, 600
geriatric psychiatric patients were transferred to other
state hospitals. A paper by Eldon Killian (1970)
reported the effects of
such transfers on the mortality rate of the patients.
\qpar
The data referring to this observation are summarized
in the table below. As already observed,  
the mean age of those who died was higher
than the mean age of the other patients. However, 
this difference can only
explain a factor 1.40 in the death rate ratio whereas
the death rate ratio with respect to those who did not
move was in fact as high as 3.95.

\begin{table}[htb]

\small

\centerline{\bf Table B1\quad Transition shock for 
relocated hospital patients. California, 1968}

\vskip 5mm
\hrule
\vskip 0.5mm
\hrule
\vskip 2mm

$$ \matrix{
\hbox{}\hfill &\hbox{Number} &\hbox{Percent}&\hbox{Average}
&\hfill\hbox{Deaths} & \hfill\hbox{Death rate}\cr
\hbox{}\hfill &\hbox{} &\hbox{of males}&\hbox{age}
&\hfill\hbox{within} & \hfill\hbox{(per year and}\cr
\qtb
\hbox{}\hfill &\hbox{} &\hbox{}&\hbox{}
&\hfill\hbox{4 months} &\hfill\hbox{per 1,000)}\cr
\noalign{\hrule}
\qth
\hbox{Patients who were moved}\hfill &144 & 61\%& 74.2 &\hfill 14 
&\hfill \hbox{\bf 292}\cr
\qtb
\hbox{Patients who did not move}\hfill & 362 & 55\%& 73.6 &\hfill 9 
&\hfill\hbox{\bf 74}\cr
\noalign{\hrule}
} $$
\vskip 1.5mm
\small
Notes: In spite of the fact that the two groups were fairly
similar in terms of age and sex proportion,
the death rate of the persons who were moved was about 4 times
higher than for those who did not move. As a matter of comparison,
in 1970 the death rate (per year and 1,000 people)
of persons aged 74 in the general
US population was comprised between 46 for females and 74 for males.
\qL
{\it Source: The data were taken from Killian (1970), 
but we lumped together
the results for 3 subgroups which, individually, were too small 
to be really significant.}
\vskip 5mm
\hrule
\vskip 0.5mm
\hrule
\end{table}


\qA{Backpedaling in the decades following 1975}

Around 1975 there was a marked turn in the US literature about
the effects of relocation on elderly people.
Whereas before the 1970s clear evidence of relocation mortality
spikes has been given in numerous papers, in the following
decades the study
of the mortality effect was dropped and research refocused
on what became known as the ``relocation stress syndrome''.
As suggested by the word ``stress'' such
studies were mostly qualitative. 
\qpar
In fact, in spite of the strong evidence
provided by earlier papers, 
the very existence of the relocation effect was put in doubt
as can be seen from the titles of the following papers:
``The relocation controversy'' (Horowitz et al. 1983),
``Relocation stress syndrome in older adults transitioning
from home to a long-term care facility: myth or reality?''
(Walker et al. 2007). 
\qpar
Did these studies really contradict earlier ones. 
If one excepts a few cases (one of which is discussed below),
they were simply looking into another direction. Thus,
the paper by Walker et al. (2007) relies on the interviews
of 16 residents of long-term care facilities that were
conducted between 2 to 10 weeks after admission. Even
with the very high death rates observed previously one would not
expect many deaths for such a small sample.
With an annual death rate of 60\%
(about 6 times the rate in the general population) the expected
number in this 8 week time interval would be 
$ 0.6\times 16\times (8/54)=1.4 $. 
In addition, the very fact that these
persons could be interviewed suggests that they were in 
good mental shape which means that they were not the most at risk.
\qpar

In the same line of thought, it can be mentioned that
some statistical data 
about nursing homes reported
in the ``Vital Statistics of the United States'' seem
doubtful. Thus, on page 379 of Vol.II, Part A of the issue of 1993,
one reads that 2,622 hospital inpatients (i.e. remaining at the
hospital overnight) committed suicide
whereas in nursing homes there were only 81. 
The total number of deaths in nursing homes was 0.4 million
which suggests that their total resident population
was about 4 millions (admitting a death rate of 100 per 1,000).
Thus the 82 suicides would correspond to an annual  suicide rate
of $ 82/40= 2.0 $ per 100,000, a rate at least 10 times lower
than in the general population of same age. 
A possible explanation was given in a book by 
Ms. Kayser-Jones (1981, p. 65):
\qdec{During the course of my field work at Pacific Manor, 
(a nursing home in California) I observed that
on several occasions when patients became terminal, they were
transferred to an acute-care hospital.
Many of these patients died shortly after transfer
from the nursing home. Thus, the number of transfers may conceal the
true number of deaths.}

\qA{No mortality in the relocation of elderly patients?
A puzzling observation}
As another illustration of the back pedaling process described
above it can be mentioned that
a study by Markson and Cumming (1974) of a transfer of 2,174 
psychiatric patients did not find any mortality spike 
in the months immediately following the transfer. 
As such a result is 
in complete contradiction with the observations described in
previous subsections, it
cannot just be reported without further
discussion (as is done in several review papers). One must 
try to understand why this observation led to a 
completely different results. There can be no science if
unexplained contradictory results are allowed to coexist. 
\qpar

This case differs from those considered previously
in several respects.
\qbu The patients were about 20 years younger than in previous
studies. Their mean age was 55. Only 7.5\% of them were over 75.
\qbu On p. 319 the authors say that patients with ``obvious
physical stigmata'' of imminent death were barred from being
transferred. However, they give no further explanations about
this selection process, for instance they do not say how many patients 
it did concern. \qL
Moreover, in two of the tables there is a fairly
mysterious category entitled ``Transferred and readmitted''.
It comprises 107 patients but not a single death 
occurred in this group during the 
11 months covered by the study%
\qfoot{The mean age in this subcategory of patients
is not given but we may 
assume that it was close to the mean age of 55 in the whole group.
Thus, based on the death rate in the general population,
one would expect 2 deaths in 11 months.}%
.
\qbu One of the authors, Elizabeth Markson, 
belonged to the team of the
``New York State Department of Mental Hygiene'' which planned
and organized the transfer. As a matter of fact, 
politically this transfer was a sensitive issue because it
was made necessary by a drastic reduction in state funding%
\qfoot{Actually, the question received nationwide attention
in May 1974 when Senator Charles Percy introduced legislation
according to which ``a decision to transfer a resident could only
be made if the transfer would be in the best interest of the patient
with full consideration of the potential impact of the change''.}%
. 
At that time, in 1974, the notion of conflict of interest
was not quite as common as it would become in subsequent decades
but from the first to the last line the paper reads as an advocacy
of the way the transfer was organized.
\qpar

These explanations do not solve the mystery completely.
The most surprising piece of evidence is the fact that in the
group of 117 women over 75 (the mean age is not given) there was not
a single death in the month following the transfer and 
only 2 in the two following months. 
If we assume a mean age of 85,
the death rate of women (in 1970) in the general population
is found to be about 10\% 
(Historical Statistics of the United States p. 62).
Thus, for 117 females over 3 months there should be
$ (3/12)\times 1.17\times 10=3 $ deaths. In other words,
instead of a mortality spike we see a death rate that is lower
than in the general population.

\qI{Appendix C. Observation of post-widowhood mortality}

In the subsection entitled ``Transient shock after widowhood''
we listed several papers in which the authors tried to measure
the mortality spike following widowhood, but we did not
summarize their results. As already mentioned, 
in such studies the main
obstacle is to collect enough data to be able to draw 
meaningful conclusions. This difficulty is explained more fully 
in the next subsection.

\qA{Rarity of statistical events for young widowers}

After an age of about 30,
according to Gompertz's law (Gompertz 1825), the human death rate
as a function of age displays an exponential increase.
Basically it is multiplied by a factor of the order of 2  
every 10 age-years. 
Thus, in 2011 in the 28 countries of the European Union, 
whereas the death rate of men was 0.90 per 1,000 in the 
age group $ 30-34 $, it reached 28.4 in the age group $ 70-74 $%
\qfoot{This results in a death rate which increases by 9\%
every year.
These data were extracted from the Eurostat database
on 20 April 2015.}%
.
\qpar
This observation
has an important implication for any investigation of the
death of widowed persons. Indeed, it shows that in young
age groups (say under 40) the joint events $ E_2 $ consisting in the
death of a spouse followed some time later by the death of the
widowed person will be very rare events.
\qpar
For instance, how many $ E_2 $ events  will there be if
for the purpose of such an investigation
one follows a group of 1 million married persons in 
the age interval $ 30-35 $
over a period of 10 years? For the sake of simplicity we can
assume that the death rate is 1 per 1,000 for married persons
and 3 per 1,000 for widowed persons.
Every year there will be about $ 10^6/1,000=10^3 $ widowed persons
and only 3 deaths of widowed persons. 
In order to be able to get a reasonable estimate of 
the distribution of the death rate 
as a function of time spent in widowhood one would need
at least 100 deaths of widowers per year.
\qpar

Event scarcity is a big obstacle. This is not only true
in the social sciences but also in the natural sciences.
For instance, the scarcity of supernovae 
in the vicinity of the solar system is a big hindrance for the
detection of gravitational waves. Similarly, the fact that
neutrinos have very few collisions with the medium through which they 
travel is a major difficulty for neutrino physics. 
\qpar

How is it possible to overcome this difficulty?\qL
One can think of four possible ways.
\qee{1} The first method which comes to mind is just to increase
the size of the age groups by studying countries with large
populations. There is another
hurdle, however, which comes from the fact that usually death
certificates do not include any information about the 
number of years that the deceased has spent in widowhood.
\qee{2} Suppose that at time $ t_0 $ there is a sudden massive and
temporary increase in the mortality rate of young people
(as for instance in October 1918  due to the influenza epidemic).
This will lead to a jump in the number of widowed persons. 
Because these persons have an inflated death rate we can hope 
to detect its effect indirectly. 

\qA{Data sources}

Basically, two kinds of sources can be used. (i) Deaths can be
recorded in a limited cohort of widowed persons followed
over several years. Examples of this kind are the
studies of Young et al. (1963), Bojanosky (1979, 1980),
Mendes de Leon et al. (1993),
Schaeffer et al. (1995),  Martikainen et al. (1996).
(ii) Deaths of widowed persons
can be identified and selected from vital statistics, more precisely
from the computer files of death certificates. Then, 
after these persons have been identified
they may be linked to their spouse through a computerized Central
Population Registry. Such a registry does exist only in some
countries, particularly in the Scandinavian countries.
\qpar

Studies which rely on vital statistics have a great advantage over
studies in which a limited cohort is followed in the course of time,
in the sense that the number of persons involved is much larger,
typically of the order of several millions instead of a few
thousands.
\qpar

In Table 2 
various studies were ranked 
according to the total number of deaths of widowed
persons involved. However, a large number of deaths is not enough
to make a good sample, age is quite as important. For a sample
of widowed persons over 80 there would of course be many deaths but
most of them would be due to old age. In other words, 
without enough young participants the 
phenomenon that we wish to observe would be largely
hidden by the Gompertz effect.
\qpar

One should also recall
that in order to determine the shape of the mortality spike
the deaths of widowed persons must be sorted by age
and by duration. This was done by Thierry (1999) for
post-widowhood durations of one or more years.
This study is summarized in the following subsection.

\qA{Response function on a yearly time scale}

To our best knowledge,
for a time resolution over one year, it is the study
by Thierry (1999) which provides the most accurate results.
This is not only due to the high numbers of events but also to the way
in which they are analyzed. Two points are of particular importance.
\qbu As age is a key variable it is crucial
to perform a 2-dimensional analysis with respect to
both age and length of widowhood. Thanks to a large number
of events, Thierry was able to distinguish as many as 6 age-groups
and 7 intervals for the length of widowhood.
\qbu It is important to present the results of the analysis
in terms of death rate ratios $ w/m $ of widowers with respect
to married persons. Why?\qL
The reason can best be explained by comparing this measure to
another one that is commonly used, namely the standardized mortality
ratio (SMR). To get the SMR the death rate of widowers is divided
by the death rate of the general population of same age.
Such a denominator is not as stable yardstick, however. For instance,
for the US population of 1970
in the female age-group 35-44, the
non-married represented 14\% whereas in the age group over 65 the
same component represented 64\%. Thus,
the SMR would fail to capture the effect that not-being married has
on death rates.

\begin{figure}[htb]
\centerline{\psfig{width=12cm,figure=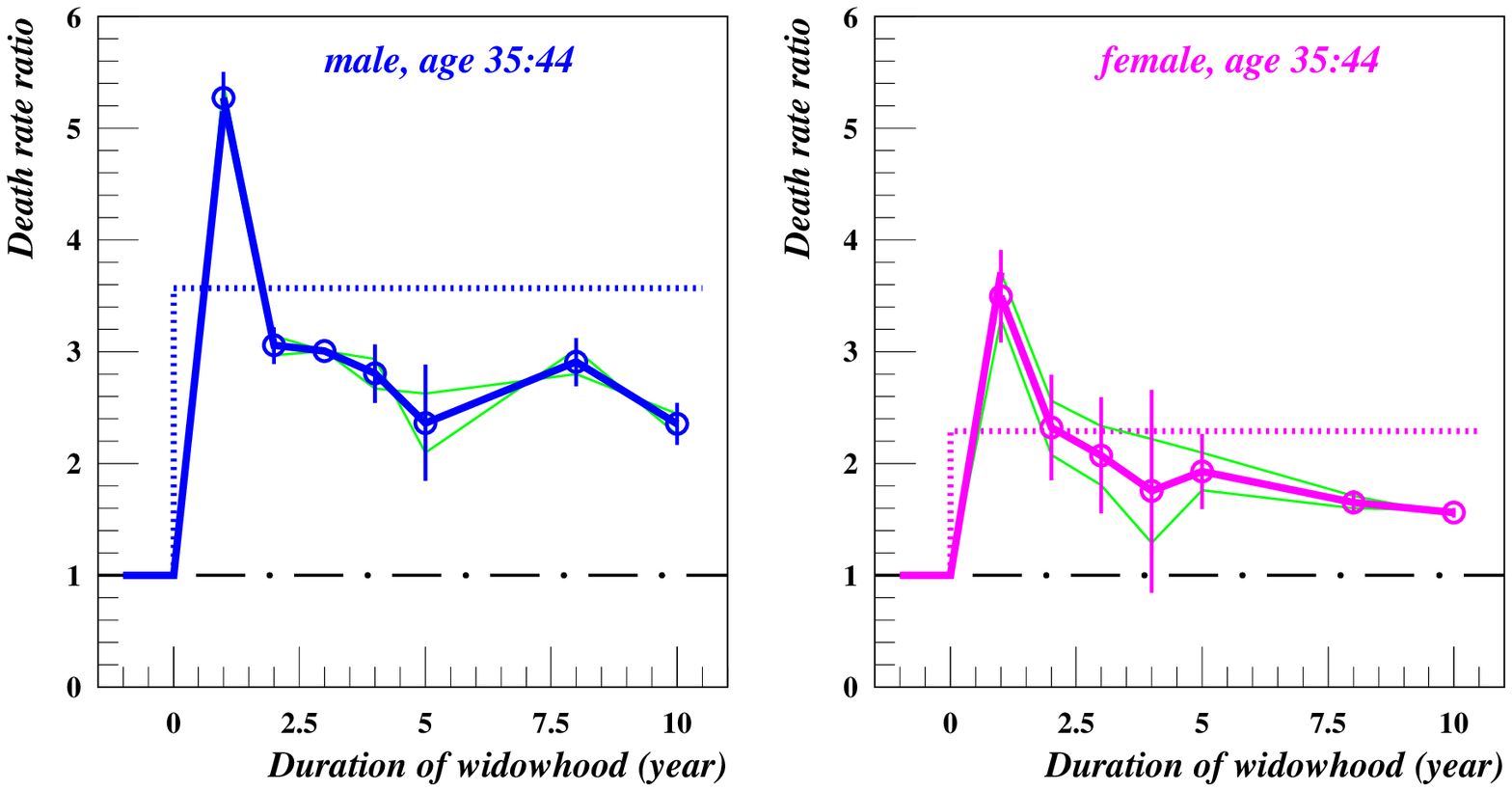}}
\vskip 3mm
\centerline{\psfig{width=12cm,figure=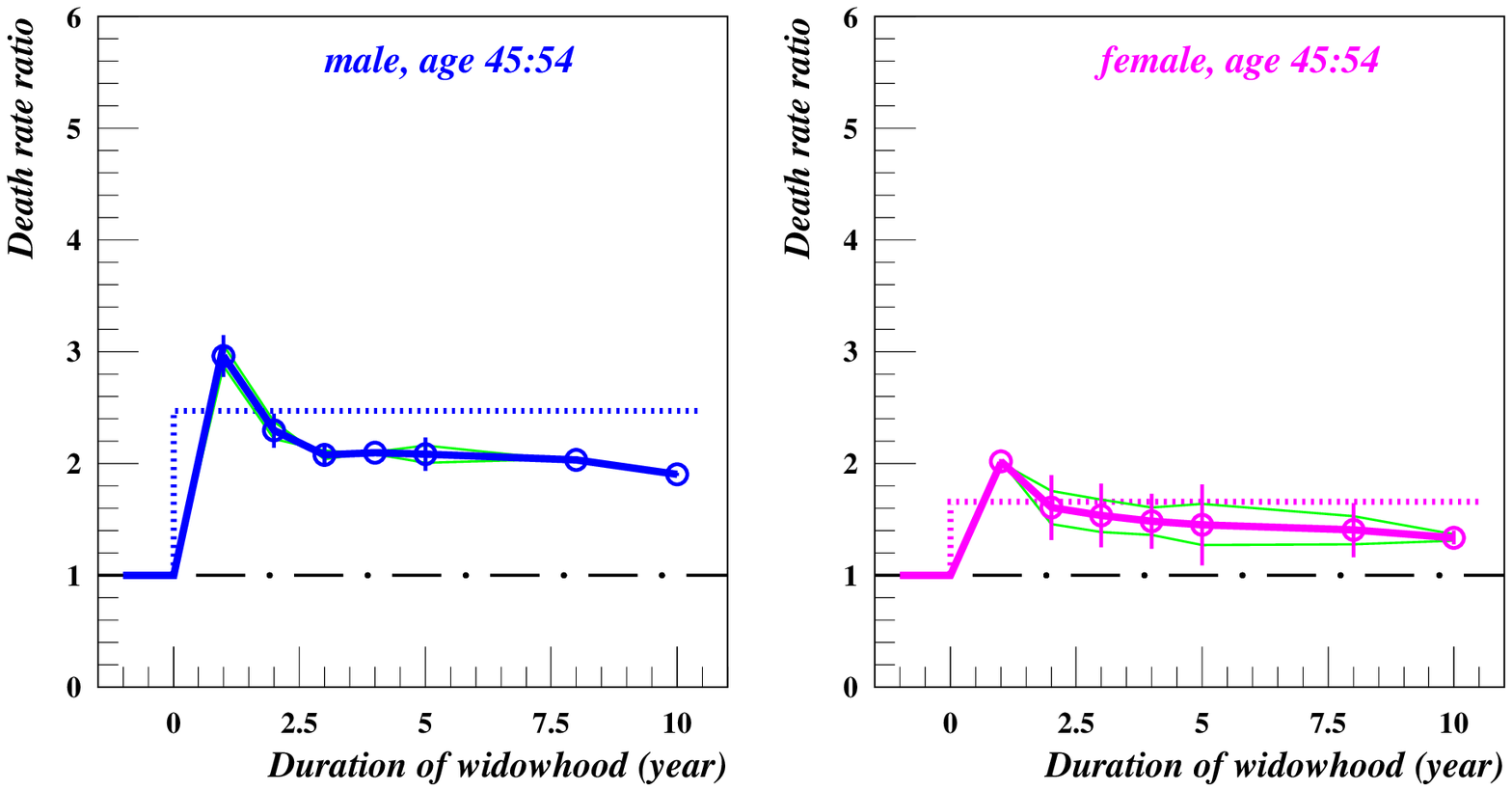}}
\vskip 3mm
\centerline{\psfig{width=12cm,figure=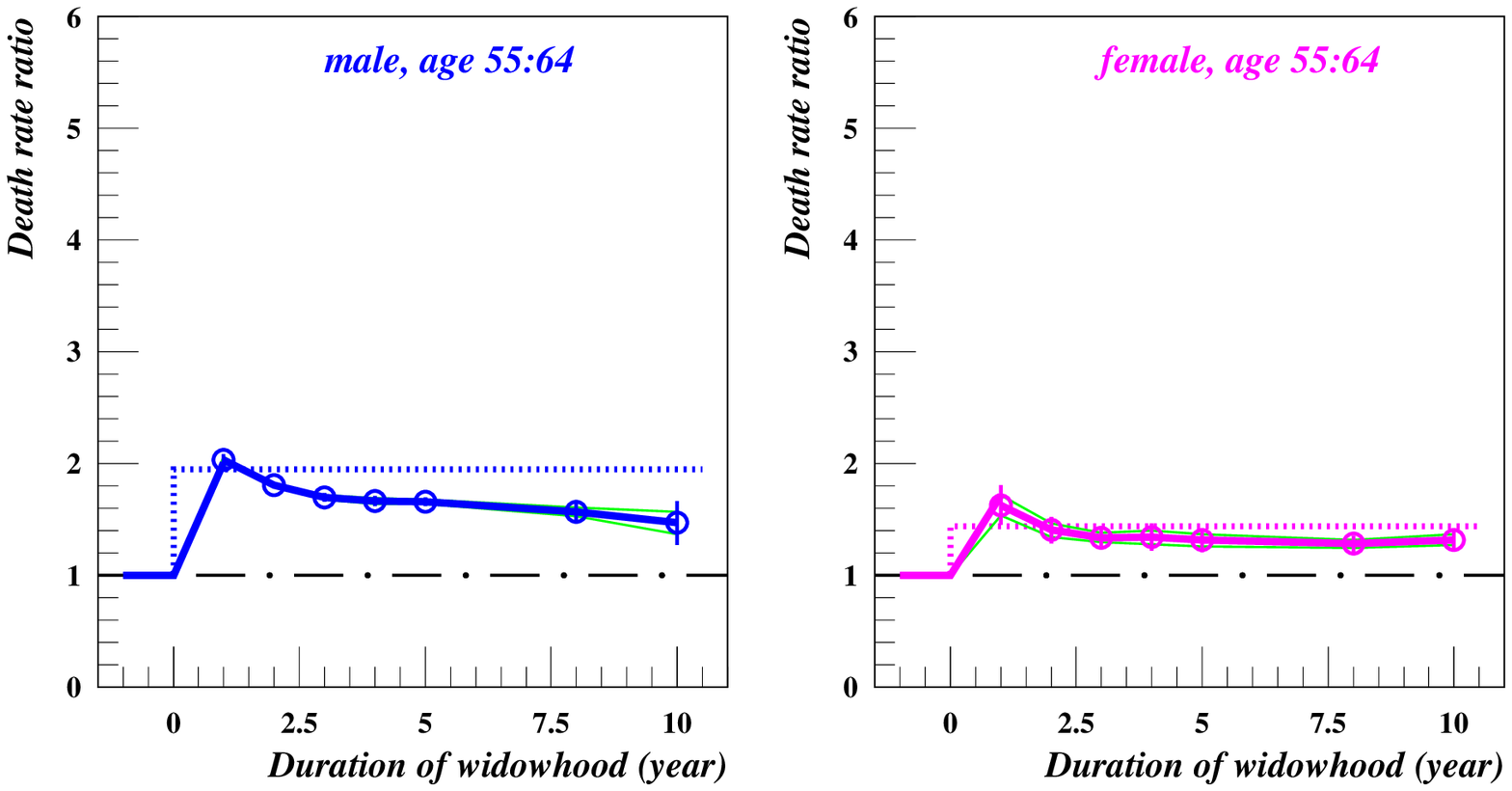}}
\qleg{Fig.\qhu C1 a,b,c\qhv Response functions of widowed persons
following death of spouse. France, 1969-1974, 1989-1991.}
{The response function gives the death rate ratio $ w/m $.
The two thin (green) lines are time series of the deaths which
occurred in the periods 1969-1974 and 1989-1991
respectively. The thick lines correspond to their averages and
the error bars are for a confidence probability level
of 0.95. As a matter of comparison, the dotted curves show
the average death ratios computed from transversal data.}
{Sources: Averages over widowhood length: 35-44:
Richmond and Roehner (2015, Table 5, USA);
45-54 and 55-54: Vallin et al. (2001, Table B and D, p. 318 and 320).
Response functions:
computations by the authors based on tables 4 and
5 of Xavier Thierry's paper (1999).}
\end{figure}

Two comments are in order regarding the graph presenting
the results obtained in Thierry's study.
\qbu The point corresponding to the first year of widowhood
is the most important because it gives the amplitude of the 
phenomenon. At the same time, it is also the one which can
obtained in the most straightforward way. Why?\qL
As already emphasized in Part 1 of this study in computing
the death rate of widowers, 
the tricky part is the determination of the
denominator that is to say the number of widowers. However,
whereas the total number of widowers is difficult to know,
the widowers who are in the first year of widowhood 
that is to say the new widowers can be counted easily through
the death certificates of the deceased spouse.
\qL
Then, in the following years, this cohort of new widowers will
shrink due to three effects: remarriage, death and emigration.
Of these factors it is emigration which is the least
documented. How important was it?
In the years covered in the graph it was not uncommon
for people who have been working in France to return to their
home country (e.g. Algeria, Portugal, Morocco or Tunisia)
after retirement. Of the three age groups that we considered,
one would expect that this phenomenon will mostly affect the oldest,
namely 55-64.  Xavier Thierry solved this problem by making
an educated guess based on data from the census of 1990.
\qbu The individual curves (thin lines) display few significant
fluctuations. As a matter of fact, the error bars are smaller
than those in the graphs of Part 1. At first sight, this may 
seem surprising because here, the analysis is more detailed
and thus necessarily relies on smaller number. One can give
two answers. \qL
The first answer is that the present analysis was
limited to persons over 35. In this way one avoids the 
youngest age groups for which the fluctuations are largest.\qL
Secondly one must recognize that the data computed
for the two time intervals (i.e. 1969-1974 and 1989-1991)
are certainly highly correlated for the simple reason that 
the same computation procedure was used. Thus, a factor like
for instance emigration was treated in the same way in the
two periods even though in reality it may have experienced
fluctuations. In other words, the error bars may underestimate
the real fluctuations.

Why did the study by Thierry (1999)
not provide a time resolution better than one year.
It seems that the 
problem came from the fact that the author got the
information about widowhood from the death certificates 
of the widowed persons (rather than by retrieving the death
certificates of the spouses as explained above)
but that only the number of years in widowhood were recorded,
not the exact date of the death of the spouse.

\qA{Is there a short-term spike immediately after bereavement?}

In the previous subsection we have seen that in the 12 months
following the death of a spouse there is an increase
in the death rate of widowed persons which is then followed
by a long phase of decrease. This shows that:
\qbu The time constant of the upgoing phase is shorter than
12 months. Needless to say, one would like to know if it
is of the order of a few days, a few weeks or a few months.
\qbu The relaxation time of the downgoing phase is of the order
of several years. 
\qpar

The observations of mortality spikes described at the beginning
of this paper
suggest similar spikes immediately after bereavement.
Is this conjecture supported by statistical evidence?

In the following subsections we give the results obtained
in three studies. 
\qee{a} The first study is for death rates by all causes; it
suggests an upgoing  time constant of the order of 6 months
and a relaxation time of the order of 2 years.
\qee{b} The second study is for suicides only; it suggests an
upgoing time constant shorter than 6 months and a relaxation
time constant of the order of 2 years.
\qee{c} The third study gives results for death by various causes.
For death by disease (e.g. heart disease) it suggests 
upgoing and downgoing time constants of a few {\it days}.
For death by suicide the time constants are the same but the
amplitude of the death rate spike is about 40 times larger.
\qpar

A key-question is whether 
these results are consistent with one another.

\qun{The Young et al. study (1963)}

The first study giving a closer insight into the timing
of the upgoing phase was the pioneering paper of Young et al. (1963).

\begin{table}[htb]

\small

\centerline{\bf Table C1\quad Response function 
of widowers in the months following bereavement.}

\vskip 5mm
\hrule
\vskip 0.5mm
\hrule
\vskip 2mm

$$ \matrix{
\hbox{Time interval following bereavement (months)} \hfill & 0-6 
& 7-12 & 13-24& 25-60  \cr
\qtb
\hbox{Duration of the time intervals (semesters)} \hfill & 1
& 1 & 2 & 6  \cr
\noalign{\hrule}
\qth
\hbox{{\bf Actual deaths of widowers} (age: 55-64, 184 deaths)} \hfill & 
  22 &  22  &  31 & 109 \cr
\hbox{{\bf Relative death rate} (per semester and per 100 deaths)}
\hfill &12 & 12  & 8.4& 9.9\cr
\qtb
\hbox{{\bf Death rate ratio} } w/m \hfill &  1.5 & 1.2 &  1.0& 1.1 \cr
\noalign{\hrule}
} $$
\vskip 1.5mm
\small
Notes: The relative death rates per semester and per 100 deaths of widowers
were computed as follows: $ (22/1)\times (100/184)=12.0 $.
It shows the distribution of deaths in the course of time
independently of their overall level.
\qL
{\it Source:  Young et al. (1963, p. 455).}
\vskip 5mm
\hrule
\vskip 0.5mm
\hrule
\end{table}


\qun{The Bojanovsky study (1979, 1980)}

\begin{figure}[htb]
\centerline{\psfig{width=16cm,figure=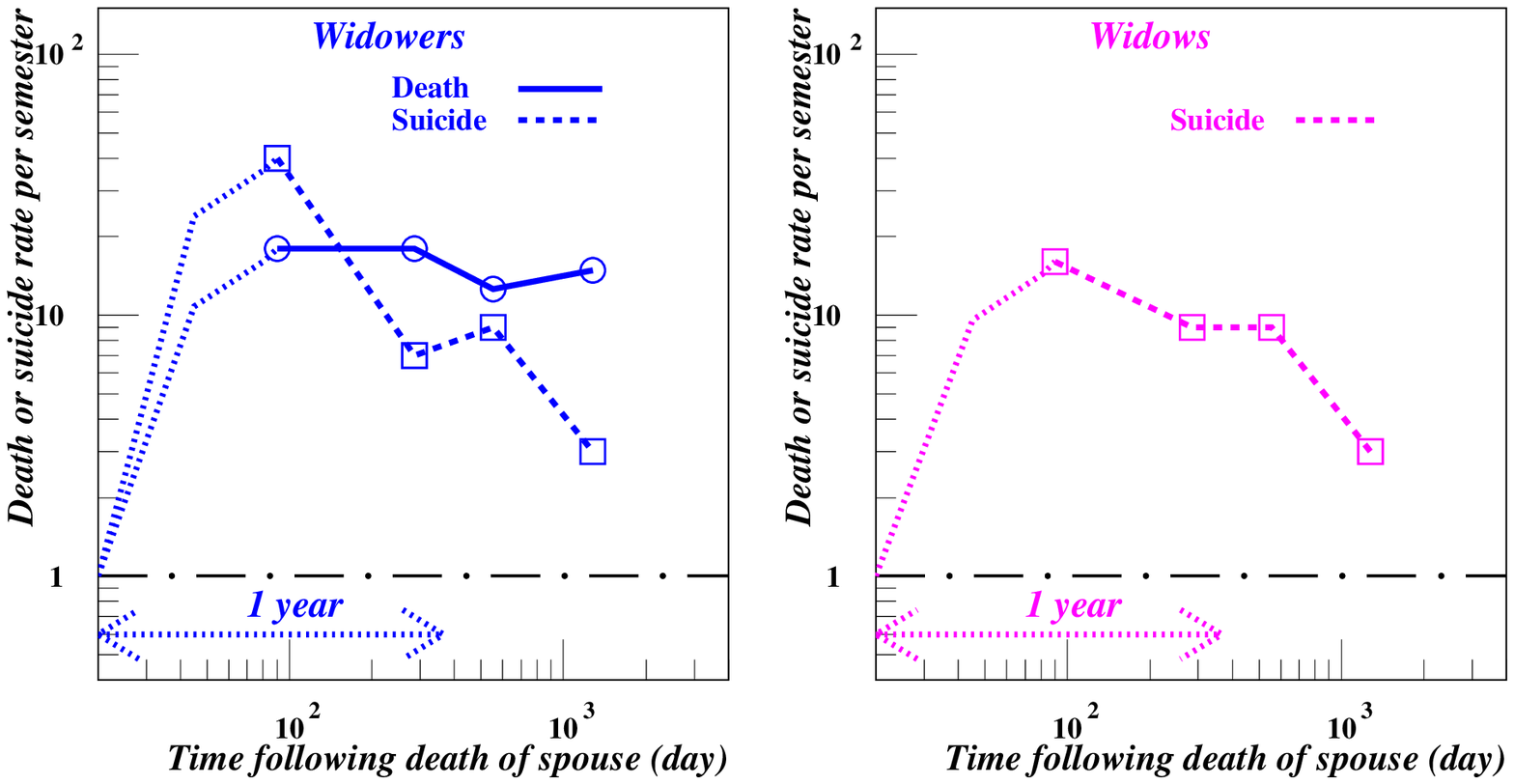}}
\qleg{Fig.\qhu C2\qhv Death response function following bereavement.}
{The graph displays the evolution of the 
rates for general mortality and
suicide following the death of  a spouse. The shortest
time interval considered was one semester.}
{Sources: Death: Young et al. (1963); suicide: Bojanovsky (1980)}
\end{figure}

 One of the clearest investigations
has been conducted by J. Bojanovsky (1979, 1980) in Germany.
He followed the following procedure.
\qbu For the cities of Heidelberg, Ludwigshaffen and Mannheim 
he was able to get police reports of suicides that occurred between
1971 and 1975. These reports gave the names and marital situation
of the suicides. 
\qbu The names of all persons who were widowed or divorced
at the time of suicide were recorded and
for almost all of them
the investigator was able to obtain the dates of the
widowhood. 
\qpar

The results are given in the table below.

\begin{table}[htb]

\small

\centerline{\bf Table C2\quad Response function to the severance
of the marital bond through widowhood or divorce.}

\vskip 5mm
\hrule
\vskip 0.5mm
\hrule
\vskip 2mm

$$ \matrix{
\hbox{Time interval following severance (month)} \hfill & 0-6 
& 7-12 & 13-24& 25-60 & 61-120 \cr
\qtb
\hbox{Duration of the time intervals (semesters)} \hfill & 1
& 1 & 2 & 6 & 10 \cr
\noalign{\hrule}
\qth
\hbox{\bf Actual suicide numbers} \hfill &  &  &  &  &  \cr
\hbox{Widowhood, males (43 cases)} \hfill & \hfill 17 & \hfill 3 
& \hfill 8 & \hfill 8 &  \hfill 7\cr
\hbox{Divorce, males (55 cases)} \hfill & \hfill 17 & \hfill 5 
& \hfill 10 & \hfill 19 & \hfill 4 \cr
\hbox{Widowhood, females (56 cases)} \hfill & \hfill 9 & \hfill 5 
&\hfill 10  & \hfill 12 & \hfill 20 \cr
\hbox{Divorce, females (27 cases)} \hfill & \hfill 1 & \hfill 1 
& \hfill 7 & \hfill 10 & \hfill 8 \cr
\hbox{} \hfill &  &  &  &  &  \cr
\hbox{{\bf Relative suicide rates per semester} (and per 100
  suicides)} \hfill &   &  &  &  &  \cr
\hbox{Widowhood, males} \hfill & \hfill 40 &\hfill 7 & \hfill 9 &\hfill 
\hfill 3 & \hfill 1.6 \cr
\hbox{Divorce, males} \hfill &\hfill  31 &\hfill  9 &\hfill  9 &\hfill
6 & \hfill 0.7 \cr
\hbox{Widowhood, females} \hfill & \hfill 16 & \hfill 9 & \hfill 9
& \hfill 3 & \hfill 3.6 \cr
\qtb
\hbox{Divorce, females} \hfill & \hfill 4 & \hfill 4  & \hfill 13 &
\hfill 6  & \hfill 3.0 \cr
} $$
\vskip 1.5mm
\small
Notes: The relative rates were computed as follows: 
$ (17/1)\times (100/43)=40 $. They show the evolution of death 
rates in the course of time and can be compared to the results
obtained by Young et al. (1963). 
It can be seen that
for widowers the rate in the first semester is about
6 times higher than in the second semester. 
It would be interesting to have detailed monthly data for the first
semester but this would require samples at least 10 times larger.
For females, the concentration of the suicides on the first
semester after dissociation is much smaller. This is not surprising
on account of the known fact that females are less affected than males
by a rupture of the marital bond.
The fairly small size of the samples in number of deceased persons
(as given within parenthesis in the first column)
shows that one cannot expect too much precision from this investigation.\qL
{\it Sources: Bojanovsky (1979, p. 75, 1980, p. 101).}

\vskip 5mm
\hrule
\vskip 0.5mm
\hrule
\end{table}


\qun{The Kaprio et al. study (1987)}

\begin{figure}[htb]
\centerline{\psfig{width=16cm,figure=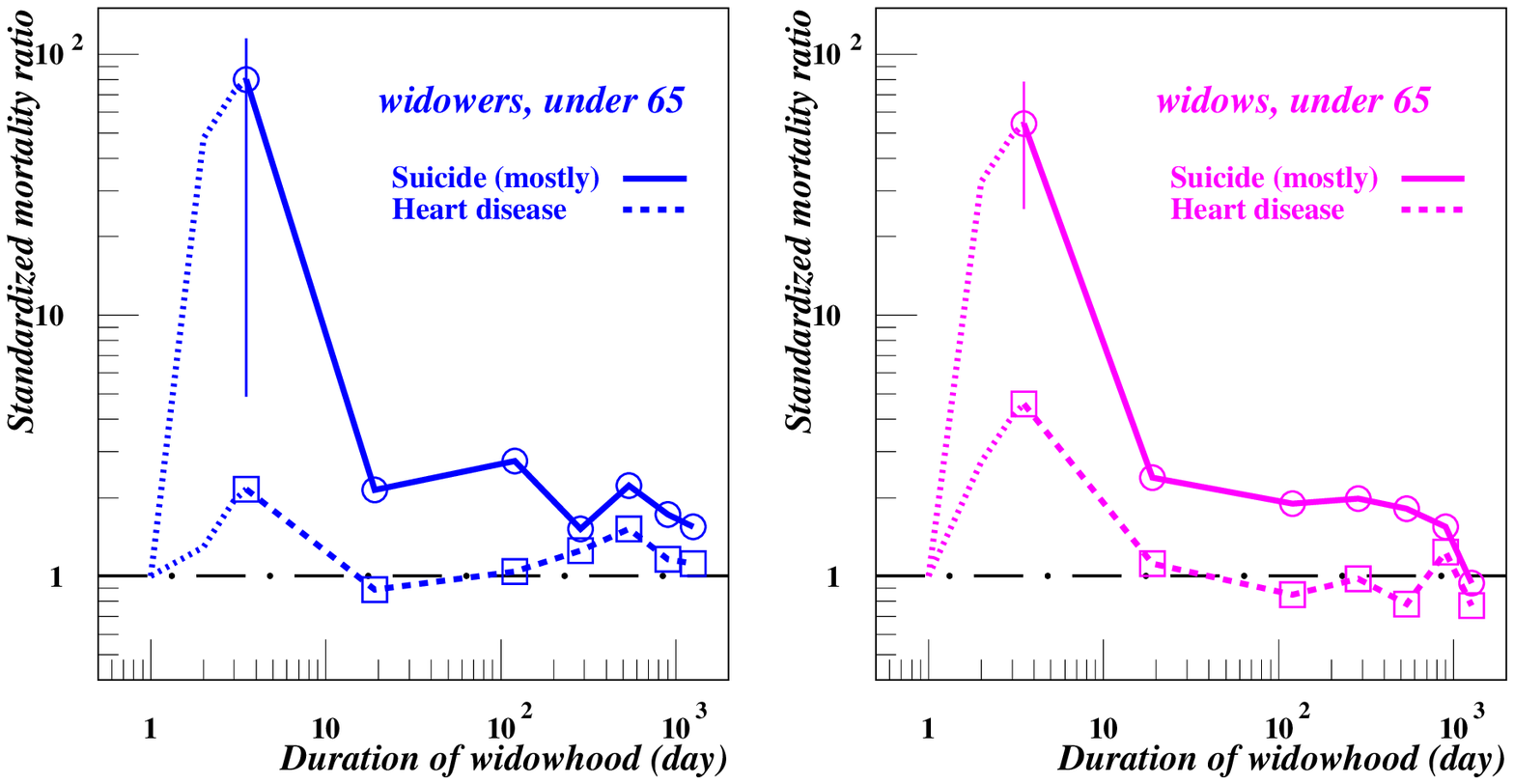}}
\qleg{Fig.\qhu C3\qhv Response function over a time scale
ranging from days to years.}
{The graph gives standard mortality ratios (SMR).
The SMR is the ratio of observed deaths in the study group
of widowed persons to deaths in the general population. 
As under age 65, most of the general population is still
married, the SMR is not very different from the death rate
ratio considered previously.
}
{Sources: The curves are based on data given in Table 4 and
Table 6 of Kaprio (1987).}
\end{figure}

This study is the only one that we could find which gives
data for time intervals after widowhood which are as short
as 1 week and 1 month. It relies on a
broad sample of 7,635 deaths of widowers which occurred between
1972 and 1976. In the Young et al. study there were only
1,540 deaths of widowers. 
\qpar

For heart disease the results show a spike 
in the first week after widowhood. For suicide%
\qfoot{Actually, the exact labeling is ``death from violent 
(non-traffic) causes'' of which suicide is the most important 
component with a weight of about 50\%.} 
there is even a  much taller spike. 
\qpar

Actually, this paper seems to contain some internal
inconsistencies%
\qfoot{For instance in the text (top of right column
on p. 284) one reads that ``exclusive of traffic accidents, male
widowers had 10 deaths from violent causes within 30 days
of the death of the spouse''. Yet, in Table 6
one reads that there were 35 non-traffic violent deaths 
of widowers in the first
week after the death of the spouse and 5 in the remaining
23 days of the first months. That means 40 deaths of that kind
in the first month.}%
.
Therefore, it would be useful to check
its results by comparing them with those
obtained by Young et al. (1963).
This is impossible however because
the authors do not give their results in the same form.
\qbu  Young et al. give total deaths whereas Kaprio et al.
give deaths by separate causes of deaths but do not
give a table for total deaths.
\qbu Young et al. give the death rate ratio with
respect to married people whereas Kaprio et al.
give the Standard Mortality Rate (SMR) that is to say
the ratio to the death rate of the total population.
\qpar
Actually, this lack of comparability is not an isolated case.
In the social sciences
comparisons are a great source of concern 
which can be described as follows.
\qpar

In physics any experimental observation is immediately
compared to similar results obtained by other researchers.
If there is a marked discrepancy, the experiment
is usually repeated until the cause of the discrepancy has been
identified. 
In the social sciences each and every paper seems to be almost 
independent from previous studies. Even when
previous studies are cited by an author their results are 
rarely discussed with the purpose to see if they are consistent
with the new results obtained by that author.
\qpar
In the present case, the paper by Kaprio et al. 
(1987) cites the study
by Young et al. (1963) but not the study by Bojanovsky (1980).
The Young et al. study is cited in the following way.
``Earlier studies on fairly small samples suggest that excess risk
is greatest immediately after bereavement (Young et al. and
other papers''. As no figures are given it is difficult to know the
real meaning of expressions such as ``greatest'' 
or ``immediately after''.  Moreover the authors do not try
to put their own results in the same form as those of Young et al.
That would be a prerequisite for any meaningful comparison.
\qpar

Social science
review papers are particularly frustrating 
as far as comparisons are concerned because
they compile and list results without describing 
the conditions under which
they were obtained and without trying to explain whether
different conditions can possibly explain observed discrepancies.

\vskip 10mm

{\bf \Large References} 

{\color{blue} 
The objective of the comments is to indicate the implications of 
those studies for the present investigation. They may be removed
in the final version.}
\qpar

\qparr
Aldrich (C.K.), Mendkoff (E.) 1963: Relocation of the aged and
disabled. A mortality study. Journal of the American
Geriatrics Society 11,3,185-194.\qL
[After the closing of the ``Chicago Home for Incurables''
its patients were relocated to other institutions.
This circumstance implies that the
relocation was not related to a deterioration of their
health.]

\qparr
Aneshensel (C.S.), Pearlin (L.I.), Levy-Storms (L.), Schuler (R.H.)
2000: The transition from home to nursing home. Mortality among
people with dementia. The Journals of Gerontology, Series B:
Psychological Sciences and Social Sciences 55,3,152-162.\qL
[The results given in the paper are based 
on the answers provided in a survey 
of 555 non-hospital caregivers who were selected because
they contacted associations concerned with dementia.
Such a source of information is certainly less reliable than the
medical records available in hospitals.
The paper presents a graph which suggests that post-admission mortality
is highest in the first month after admission
even for persons who were not
admitted for compelling health reasons. Unfortunately,
not only does the graph not show any error bars but even the data points
themselves are not displayed.]

\qparr
Beach (B.) 2013: How long do disk drives last?\qL
The website:
https://www.backblaze.com/blog/how-long-do-disk-drives-last/
provides data covering a 4-year period at the cloud storage
company Backblaze.

\qparr
Bertillon (J.) 1879: Les c\'elibataires, les veufs et les
divorc\'es du point de vue du mariage. [Attitude with respect
to marriage of non-married, widowed and divorced persons.]
Revue Scientifique de la France et de l'Etranger 8,33,776-783.

\qparr
Bertillon (L.-A.) 1872: Article ``Mariage'' in the
Dictionnaire Encyclop\'edique des Sciences M\'edicales,
[Encyclopedic Dictionary of the Medical Sciences].
2nd series, Vol. 5, p.7-52.

\qparr
Bertillon (L.-A.) 1879: Article ``France'' in the
Dictionnaire Encyclop\'edique des Sciences M\'edicales,
[Encyclopedic Dictionary of the Medical Sciences].
4th series, Vol. 5, p.403-584.

\qparr
Bojanovsky (J.) 1979: Wann droht der Selbstmord bei Geschiedenen?
[After a divorce when is the likelihood of committing 
suicide largest?].
Schweizer Archiv f\"ur Neurologie, Neurochirurgie und Psychiatrie
125,1,73-78.

\qparr
Bojanovsky (J.) 1980: Wann droht der Selbstmord bei Verwitweten?
[After becoming a widower when is the likelihood of committing
suicide largest?].
Schweizer Archiv f\"ur Neurologie, Neurochirurgie und Psychiatrie
127,1,99-103.

\qparr
Boyle (P.J.), Feng (Z.), Raab (G.M.) 2011: Does widowhood increase
mortality risk? Testing for selection effects by comparing
causes of spousal death.
Epidemiology 22,1,1-5.\qL
[The study relies on a data set from the Scottish Longitudinal
Study which covers 5\% of the Scottish population from 1991 to 2001.
A total of 14,630 persons were widowed in the sample studied by the
authors. Throughout their paper the authors use the notion of
hazard ratio which generalizes the notion of death ratio 
(used in Part I of our study) to 
various cases of risks, e.g. the risk of non fatal heart attack. When
specifically focused on the risk of death, the hazard ratio is identical
to the death ratio. The authors found a death ratio widowed/married
of $ 1.40 \pm 0.07 $ which is consistent  with the results obtained
by Young and al. (1965) but, in contrast to them, they found almost
no duration effect that is to say no higher mortality shortly
after widowhood.]

\qparr
Camargo (O.), Preston (G.H.) 1945: What happens to patients
who are hospitalized for the first time when over sixty-five years
of age. American Journal of Psychiatry 102,2,168-173.\qL
[The title is somewhat misleading. These patients were not
hospitalized temporarily. They were admitted to mental hospitals
from where less than 10\% were discharged in subsequent years.
The authors gave special attention to the mortality rate in the
months following admission. They found that approximately
16\% of all admitted patients died during the first month after
admission and 46\% had died by the end of the first year.]

\qparr
Durkheim (E.) 1897: Le suicide. Etude de sociologie. F. Alcan, Paris.
A recent English translation is: ``On Suicide'' (2006),
Penguin Books, London.\qL
[Durkheim showed not only that 
suicide rates among non-married, widowed or divorced
persons were higher than among married persons but also
that they were lower among married persons with several
children than for married persons without children.]

\qparr
Farr (W.) 1859, 1975: Influence of marriage on the mortality of the
French people (12 p.). Transactions of the National Association
for the Promotion of Social Science 1858-1859, 504-520.
The paper was republished in 1975 in ``Vital statistics, a memorial
volume of selections from reports and writings of William Farr''.
Scarecrow Press, Methuen (New York).
\qL
[The style of the report is somewhat outdated and confusing.
An illustration is provided
by the following excerpt taken from the first paragraph.
``The action of the various parts of the body in industrial occupations
  produces specific effects. Every science modifies its
  cultivators. The play of the passions transfigures the human
  frame. How do they influence its existence?'']

\qparr
Frisch (M.), Simonsen (J.) 2013: Marriage, cohabitation and mortality
in Denmark: national cohort study of 6.5 million persons followed
for up to 3 decades. International Journal of Epidemiology 1,13.

\qparr
Gompertz (B.) 1825: On the nature of the function expressive of
the law of human mortality, and on a new mode of determining the value
of life contingencies. Philosophical Transactions of the Royal
Society 115,513–585. 

\qparr
Gove (W.R.) 1972: Sex, marital status, and suicide. Journal of
Health and Social Behavior 13,204-213.

\qparr
Grove (R.D.), Hetzel (A.M.) 1968:  Vital statistics rates 
in the United States, 1940-1960. United States Printing Office,
Washington DC. 

\qparr
Hayes (L.M.), Rowan (J.R.) 1988: National study of jail suicides:
seven years later. National Center on Institutions and
Alternatives, Alexandria (Virginia).

\qparr
Helsing (K.J.), Szklo (M.), Comstock (G.W.) 1981: Factors 
associated with mortality after widowhood. 
American Journal of Public Health 71,802-809.\qL
[The paper is based on the comparison of a sample of 4,032 persons
who became widowed between 1963 and 1974 and a sample
of same size of married persons. For males, the ratio of death
rates of widowed persons to that of married persons is 1.34.
The paper is freely available on the Internet.]

\qparr
Horowitz (M.S.R.), Schultz (R.) 1983: The relocation controversy.
Criticism and commentary on 5 recent studies. 
The Gerontologist 21,229-234.

\qparr
IPUMS: Integrated Public Use Microdata Series, 
University of Minnesota, Minneapolis.\qL
[The IPUMS website offers a set of databases giving individual census 
records. It covers
the United States as well as a number of other countries.]

\qparr
Kaprio (J.), Koskenvuo (M.), Rita (H.) 1987: Mortality
after bereavement. A prospective study of 95,647 widowed persons.
American Journal of Public Health 77,3,283-287.\qL
[The study involved the deaths of 7,635 widowed persons.
The breaking up by age groups was limited to only two
groups: under versus over 65. Like the Martikainen paper,
the authors are 
are rather interested in the distinction between various causes
of death.]

\qparr
Kayser-Jones (J.S.) 1981: 
Old, alone, and neglected.
Care of the aged in the United States and Scotland.
University of California Press, Berkeley.\qL
[After staying 3 months in a nursing home in the east of Scotland
and 4 months in a private nursing home in California, the author
provides a comparative analysis.]

\qparr
Killian (E.C.) 1970: Effect of geriatric transfers on mortality rates.
Social Work, 15,1,19-26.\qL
[The author followed for 4 months some 144 patients who were
transferred from the Stockton hospital in California to
other facilities. Actually, 387 patients were transferred
between January 1966 and March 1968.
The author does not say why he limited his study to the 144 who
were relocated between January and March 1968. 

\qparr
Krose (H.A.) 1906: Der Selbstmord im 19. Jahrhundert. 
[Suicide in the 19th century].
Herdersche Verlag, Fribourg.

\qparr
Linder (F.E.), Grove (R.D.) 1947: Vital statistics rates 
in the United States, 1900-1940. United States Printing Office,
Washington DC. 

\qparr
Locoh (T.) 1972: L'entr\'ee en maison de retraite. Etude aupr\`es
d'\'etablissements de la r\'egion parisienne.
[Admission into nursing homes. A study of institutions located
in Paris and the surrounding area.]
Population 27,6,1019-1044.\qL
[The author followed a population of about 600 persons (125 men
and 480 women) following their admission into nursing
homes.]

\qparr
March (L.) 1912: Some researches concerning the factors of 
mortality. Journal of the Royal Statistical Society 75,505-538.\qL
[The paper includes a comparative analysis of mortality rates
in France, Prussia and Sweden.]

\qparr
Markson (E.W.), Cumming (J.H.) 1974: A strategy of necessary mass
transfer and its impact on patient mortality.
Journal of Gerontology 29,3,315-321.\qL
[The paper relates the transfer between different hospitals
of 2,174 psychiatric patients in New York State. 
There were two major differences
between this study and similar ones. Firstly, the study did
not find any mortality spike in the weeks following the transfer.
Secondly, one of the authors, Elizabeth Markson, 
belonged to the team of the
``New York State Department of Mental Hygiene'' which planned
and organized the transfer.  A more detailed discussion 
is given in the paper.]

\qparr
Martikainen (P.), Valkonen (T.) 1996: Mortality after the 
death of spouse. Rates and causes of death in a large Finnish
cohort. American Journal of Public Health 86,8,1087-1093.\qL
[This study relied on a total of 9,935 deaths of widowed persons
which is a substantial
number but, in contrast with most other studies, it
found only a small difference in death ratio during
the first 6 months after widowhood and the subsequent 5 years,
namely 1.29 compared with 1.19.
One may wonder why? The reason is fairly simple. 
The authors did not make an analysis by age groups but instead
lumped all ages together. As the sample contained people
who were between 40 and 89 at the end of the study, the results
are completely dominated by the deaths of old persons, say over 75.
It is not surprising that in such an age group, the mortality ratio is 
of the order of 1.3 and that it decreases very little in
the course of time. One can regret that the authors did not
extract from their study relevant data for younger age groups.]

\qparr
Mellstr\"om (D.), Nilsson (A.), Od\'en (A.), Rundgren (A.), 
Svanborg (A.) 1982: Mortality among the widowed in Sweden.
Scandinavian Journal of Social Medicine 10,33-41.\qL
[The study followed
360,000 individuals aged between 50 and 90
who were widowed in Sweden at some point between 1968 and 1978. 
It showed a peak in the mortality risk during the first 3 months.
For widowers the amplitude (with respect to married persons) was 1.48
while for widows there was a peak of smaller amplitude, namely 1.22.]

\qparr
Mendes de Leon (C.), Kasi (S.), Jacobs (S.) 1993: Widowhood and
mortality risk in a community sample of the elderly. A prospective
study. Journal of Clinical Epidemiology 46,6,519-527.\qL
[The study involved only 22 deaths of widowed persons.]

\qparr
Mortality detail file for 1992 (ICPSR 6798) 2001: Published 
by the US National 
Center for Health Statistics. \qL
[It is the user guide of an electronic file of {\it all} the
deaths that occurred in the United States in 1992. It gives the 
meaning of the codes of all variables. The same kind
of data is available on the website of
the ``Consortium for Political and Social
Research'' for all years from 1968 to 1992.]

\qparr
Mortality statistics: review of the Registrar General on deaths
in England and Wales. Series DHI, Number 16. Her Majesty's Stationary
Office, London. 

\qparr
Mortality Statistics 1910 (published in 1912): Bulletin 109. Bureau
of the Census, Government Printing Office, Washington DC.

\qparr
Nagle (J.T.) 1882: Suicides in New York City during the 11 years
ending Dec. 31, 1880. Riverside Press. Cambridge (Ma.).

\qparr
National Center for Health Statistics (NCHS) 1970: Mortality
from selected causes by marital status. Vital and Health
Statistics, Series 20, number 8.

\qparr
Parkes (C.M.), Benjamin (B.), Fitzerald (R.G.) 1969:
Broken heart. A statistical study of increased mortality among
widowers. British Medical Journal 1,740-743.
\qL
[This is the continuation of the Young et al. (1963) study.
The same sample was observed over 4 more years i.e. a total
of 9 years after the death of the spouse. Over these 4 years
the death ratio widowed/married was comprised between 0.90 and
0.95.]

\qparr
Registrar General 1971: Statistical Review of England and Wales.
Part III. Office of Population Censuses and Surveys. London.

\qparr
Richmond (P.), Roehner (B.M.) 2015: Effect of marital status on
death rates. Part 1: High accuracy exploration of the 
Farr-Bertillon effect.
Preprint, May 2015.

\qparr 
Roehner (B.M.) 2007: Driving forces in physical, biological
and socio-economic phenomena. Cambridge University Press,
Cambridge.

\qparr
Sattar (G.) 2001: Rates and causes of death among prisoners 
and offenders under community supervision. Home Office Research Study
231.

\qparr
Schaeffer (C.), Quesenberry (C.), Wi (S.) 1995: Mortality following
conjugal bereavement and the effects of a shared environment.
American Journal of Epidemiology 141,12,1142-1152.\qL
[The study involved 934 the deaths of 934 widowed persons.]

\qparr
Statistisches Jahrbuch f\"ur die Bundesrepublik Deutschland 1978:
Stuttgart.

\qparr
Stroebe (W.), Stroebe (M.S.) 1987: Bereavement and health.
The psychological and physical consequences of partner loss.
Cambridge University Press, Cambridge.\qL
[The book is mostly concerned with psychological and
other qualitative aspects. Although short (p. 151-167), the
review of the quantitative evidence about the mortality of widowed
persons is quite useful. It can be noted that the book has
a broad reference section which contains about 600 entries.]

\qparr
Thierry (X.) 1999: Risques de mortalit\'e et de surmortalit\'e au
cours des 10 premi\`eres ann\'ees de veuvage.
[Excess mortality during the first 10 years of widowhood.] 
Population 54,2,177-204. \qL
[This is a key-paper in the investigation
of the impact of widowhood duration
on the death rate of widowers. It is not based on a 
sample but on the whole French population followed over 
8 years.
It takes advantage of the fact that, in contrast with most
other countries,  French death certificates contain
information about widowhood length.

\qparr
Thierry (X.) 2000: Risques de mortalit\'e et causes m\'edicales
des d\'ec\`es aux divers moments du veuvage. [Mortality and causes
of death throughout widowhood]
G\'erontologie et Soci\'et\'e 95,27-43.

\qparr
Vallin (J.), Mesl\'e (F.), Valkonen (T.) 2001: 
Tendances en mati\`ere de mortalit\'e et mortalit\'e diff\'erentielle.
Editions du Conseil de l'Europe, Strasbourg.\qL
An English version was published under the title
``Trends in mortality and differential mortality''. 

\qparr
Walker (C.A.), Curry (L.C.), Hogstel (M.O.) 2007: Relocation stress
syndrome in older adults transitioning from home to
a long-term care facility: myth or reality?
Journal of Psychological Nursing and Mental Health Services 45,1,38-45.

\qparr
Wang (L.), Xu (Y.), Di (Z.), Roehner (B.M.) 2013:
How does group interaction and its severance affect life expectancy? 
arXiv preprint 1304.2935 (9 April 2013) 

\qparr
Weyerer (S.), Wiedenmann (A.) 1995: Economic factors and the
rates of suicide in Germany between 1881 and 1989. 
Psychological Reports 76,1331-341.

\qparr
Wilkins (D.J.) 2002: The bathtub curve and product failure behavior.
Part I: The bathtub curve, infant mortality and burn-in.
Hot Wire 21 (November 2002)

\qparr
Young (M.), Benjamin (B.), Wallis (C.) 1963: Mortality of
widowers. Lancet 2,254-256.\qL
[This is a longitudinal study. It involved the deaths
of 906 widowers.
The authors followed over a period of 5 years
a sample of 4,486 widowers more than 55 years old and
whose wives had died in January 1957. Altogether there
were 906 deaths of widowers.
In the age-group 60-64 they found a 
death rate ratio widowed/married of 1.7 in the first six months after
the death of the spouse and 1.0 for the remaining time;
however, it must be observed that there were only 44 deaths 
in this age-group.\qL
The problem
is that these ratios are too low to account for the ratio
of 1.6 given by UK vital statistics data for 
age groups which cover 5 or 10 years. On average individuals
spend 5 years in a 10 year age group. If their death rate is 
multiplied by $ k_1 $ over an interval of $ m $ months, 
and by $ k_2 $ for the remaining months, then
for the 5 years during which they remain in the age group
the average rate will be multiplied by:
 $ k=m\times k_1 + 54\times k_2)/60 $;
for $ m=6, k_1=1.39, k_2=1.04 $ one gets: $ k=1.08 $.]

\end{document}